\documentclass[11pt,a4paper]{article}
\usepackage[utf8x]{inputenc}
\usepackage[T1]{fontenc}
\usepackage{mathptmx} 

\usepackage[margin=25truemm]{geometry}
\usepackage[pdftex]{graphicx} 
\usepackage{calc} 
\usepackage{authblk}
\usepackage{booktabs}
\usepackage{bm}
\usepackage{enumitem} 
\usepackage{braket,mathtools,amsmath,amssymb,subcaption}
\usepackage{authblk}
\usepackage[all]{nowidow} 
\usepackage{url}
\usepackage[normalem]{ulem}

\usepackage{lipsum} 

\usepackage{framed} 
\usepackage[framed]{ntheorem}
\newframedtheorem{frm-thm}{Theorem}

\def\a{\alpha}

\def\g{\eta}

\def\hx{\hat{x}}
\def\hy{\hat{y}}
\def\hz{\hat{z}}
\def\hhx{\hat{\mathbf x}}
\def\hhy{\hat{\mathbf y}}

\def\br{\mathbf{r}}

\usepackage{hyperref}
\hypersetup{colorlinks,linkcolor=blue,citecolor=blue}

\makeatletter
\def\ps@preprint{%
  \def\@oddhead{\hfill\small RIKEN-iTHEMS-Report-26, STUPP-26-307}%
  \let\@evenhead\@oddhead
  \def\@oddfoot{}%
  \let\@evenfoot\@oddfoot
}
\makeatother
\begin{document}

\title{Statistical mechanics and hierarchy of learnability 

in quantum codes: From generalized symmetries to non-Abelian topological orders}

\author{Hiromi Ebisu}
\author{Masazumi Honda}
\affil{Interdisciplinary Theoretical and Mathematical Sciences Program (iTHEMS)~RIKEN, Wako 351-0198, Japan}

\affil{Graduate School of Science and Engineering, Saitama University, 255 Shimo-Okubo, Sakura-ku, Saitama 338-8570, Japan}

\maketitle
\thispagestyle{preprint}

\begin{abstract}
Local weak measurements provide noisy information about observables encoded in a quantum code. We formulate the corresponding Bayesian inference problem for a broad class of topological stabilizer codes by mapping it to classical spin models with generalized symmetries. When the wavefunction deformation and Gaussian measurement strengths coincide, the replicated model acquires an enlarged symmetry. We use this symmetry to derive exact identities among correlators averaged over measurement records that quantify the learnability of nonlocal observables associated with fractionalized excitations. Together with correlation inequalities in the associated clean ferromagnetic models, these identities establish hierarchies of learnability: if two observables remain learnable at long distances, their product must also remain learnable. We demonstrate these relations in the rank-2 toric code and the X-cube model, whose classical descriptions exhibit modulated and subsystem symmetries, respectively. We further extend the analysis to finite group quantum double models, including non-Abelian ones. We derive exact relations between learnability and correlations in the corresponding classical models, and show that the representation structure and fusion rules of anyons impose further constraints under suitable positivity conditions. These results constrain which types of nonlocal information can be learned simultaneously, providing a guide to learning phase diagrams in quantum codes under local weak measurements.
\end{abstract}

\newpage
\pagenumbering{arabic}

\tableofcontents
\section{Introduction}

Topological quantum error correcting codes store information in nonlocal
degrees of freedom that are protected against local perturbations
\cite{KITAEV20032,dennis2002topological,PhysRevLett.103.090501}.
Local weak measurements provide partial and noisy information about the quantum state~\cite{PhysRevLett.129.120604,7dpt-d4s5,Putz:2026learning}.
This raises a natural inference problem: “given such a measurement record, which
nonlocal observables can be inferred, and what constraints relate their
learnability?” For a binary observable in a stabilizer code, a natural estimate is its average conditioned on the measurement record, called the posterior expectation value. Its square, averaged over all records, diagnoses how reliably the observable can be inferred. The behavior of this quantity at large separations therefore indicates whether the corresponding nonlocal information remains learnable over long distances.

Statistical mechanical mappings provide a powerful framework for
addressing this problem.
They describe both Pauli errors in stabilizer codes~\cite{dennis2002topological,PhysRevB.110.085158,chubb2021statistical,Lyons:2024mapping} and the inference problem associated with weak measurements ~\cite{Nishimori_1980,nishimori2001statistical,iba1999nishimori,PhysRevLett.61.625,PhysRevB.40.9249,KovalevPryadko2013,zdeborova2016statistical,b3s8-y1wb}. In the latter case, the spin configuration is unknown, and Bayes’ theorem uses the noisy measurement record to update the probability assigned to each configuration. 
Recent studies of the Ising learning problem and its dual description in terms of a deformed toric code wavefunction~\cite{PhysRevB.77.054433} identified a special relation between the model parameters, called the \textit{higher Nishimori line}~\cite{patil2026higher,Wiese:2026bayesian}.
For Gaussian weak measurements, this line is obtained by setting the measurement strength equal to the strength of the wavefunction deformation. This condition leads to an enlarged permutation symmetry in a formulation using several copies, or replicas, of the classical model, allowing us to obtain exact relations between various record averaged correlators. Related higher Nishimori structures have recently been identified in
Potts models and deformed $\mathbb{Z}_q$ toric codes
\cite{Patil:2026potts}.

These results motivate a further question. Previous studies have mainly related averages of different powers of the posterior expectation of a single observable. In quantum codes with generalized symmetries, distinct nonlocal observables can also obey nontrivial multiplication relations. We ask whether the higher Nishimori symmetry can turn these relations into exact constraints on which observables can be learned simultaneously.

In this work, we address this question for a broad class of
Calderbank--Shor--Steane (CSS) stabilizer codes and extend the analysis
to finite group quantum double models, including non-Abelian
ones.
For CSS codes, the commutation relations between stabilizer generators
and locally measured Pauli operators directly determine the interactions
of an associated classical spin model.
Each measured Pauli operator produces a local multispin interaction,
while relations among the stabilizer generators appear as exact
symmetries of the classical model.
This construction accommodates ordinary global, spatially modulated,
and subsystem symmetries within a common framework.\par
For these codes, matching each wavefunction deformation strength to the
corresponding Gaussian measurement strength produces an enlarged
replica permutation symmetry.
We refer to the resulting parameter manifold as the
\textit{higher Nishimori manifold}.
Using this symmetry, we derive exact identities relating the posterior
expectations of distinct observables and their operator products.
In particular, the record averaged second moments that quantify
learnability are equal to correlations in the corresponding clean
models. Here, the clean model describes the deformed state before conditioning on a measurement record.
Combining this relation with correlation inequalities for clean models
with nonnegative ferromagnetic couplings, we establish hierarchies of
learnability: if two observables remain learnable at long distances,
their product must also remain learnable.
These inequalities constrain the possible patterns of learned and
unlearned information without requiring a complete solution of the
measurement dependent phase diagram.

As a concrete application, we consider the rank-2 toric code
\cite{PhysRevB.106.045145}.
The associated classical model possesses spatially modulated symmetries
inherited from relations among the rank-2 toric code stabilizers
\cite{Ebisu:2023idd,Pace:2025hpb,10.21468/SciPostPhys.20.4.117}.
Correspondingly, the quantum code supports spatially modulated
dipole loop operators.
Multiplying a dipole loop operator by a lattice translation gives a homogeneous monopole loop
operator.
Applying the general hierarchy to the corresponding open string
observables, we show that long range
dipole learnability implies long range monopole learnability.
Hence, the multiplication relations among these operators directly
restrict the allowed patterns of learnability.\par
We obtain analogous exact relations in the X-cube model
\cite{Vijay,johnston2017plaquette,PhysRevLett.129.230502}.
For local Pauli-$Z$ measurements, relations among cube stabilizers
generate planar subsystem symmetries in the associated classical
plaquette model.
Multiplication relations among subsystem invariant observables then
yield exact inequalities among learning correlators that probe
different pairs of planes.

Beyond CSS stabilizer codes, we extend the analysis to finite group quantum double models~\cite{KITAEV20032,PhysRevB.78.115421,Wootton:2010dah,Cui:2019lvb,bravyi2022adaptive},
including the ones with non-Abelian anyons.
We
establish an exact higher Nishimori identity for an isotropic Gaussian
measurement protocol.
This identity relates the learning correlators of pure electric
observables to correlations in the corresponding classical model.
When the classical interactions admit a nonnegative expansion in terms of normal subgroups, we derive
further inequalities.
We illustrate these constraints in the~$S_3$ and $D_4$ quantum doubles.
Together, these examples demonstrate how operator multiplication and
fusion constrain which types of nonlocal information
can be learned simultaneously from noisy local measurements, providing a guide to learning phase diagrams in quantum codes.

The remainder of this paper is organized as follows.
In Sec.~\ref{sec:classical-mapping}, we review the classical spin
representation of CSS stabilizer codes, including the symmetries
arising from relations among stabilizer generators, and formulate the
weak measurement problem as Bayesian inference.
In Sec.~\ref{app:generalized-HN}, we derive the generalized higher
Nishimori identities and the resulting hierarchy of learnability.
We apply these results to the rank-2 toric code in Sec.~\ref{exx},
with the X-cube analysis presented in Appendix.
In Sec.~\ref{QD}, we formulate the learning problem for finite group
quantum doubles and derive constraints on the learnability of their
electric observables.
Finally, Sec.~\ref{oput} summarizes our results and discusses future directions.
Additional technical details are provided in the appendices.

\section{From stabilizer codes to classical spin models}
\label{sec:classical-mapping}
Before formulating the learning problem, we review how a CSS
stabilizer code gives rise to a classical spin model.  We focus on
one Pauli sector, since the two sectors of a CSS code can be
treated independently in the same manner.  Previous works have
established the correspondence between quantum information
observables of stabilizer codes and observables of the
associated classical spin models~\cite{dennis2002topological,PhysRevB.110.085158}.  Our purpose here is to
formulate the measurement induced inference problem within this
correspondence and to identify the role played by stabilizer
relations.
\subsection{Stabilizer relations and spin symmetries}\label{sec21}
Let $\{S_k\}_{k=1}^{N_s}$ be a set of mutually commuting
stabilizer generators, all of either $X$ type or $Z$ type.  An
element of the subgroup generated by these operators can be
written as
\begin{equation}
    g(\bm q)
    =
    \prod_{k=1}^{N_s}S_k^{q_k},
    \qquad
    q_k\in\mathbb F_2.\label{1}
\end{equation}
Here $\mathbb F_2=\{0,1\}$ is the binary field, with addition
and multiplication defined modulo two.
We associate an Ising variable with each binary variable $q_k$
according to
\begin{equation}
    \sigma_k=(-1)^{q_k},
    \qquad
    q_k=\frac{1-\sigma_k}{2}.
\end{equation}
Thus, $q_k=1$, corresponding to the inclusion of $S_k$ in
$g(\bm q)$, is represented by $\sigma_k=-1$.
We next introduce a family of local Pauli operators $\{P_j\}$
of the complementary Pauli type.  These operators represent
local errors in Sec.~\ref{subsec:classical-weight} and the observables
subjected to weak measurements in
Sec.~\ref{sc3}.  Their commutation relations
with the stabilizer generators define the binary commutation
matrix $A$:
\begin{equation}
    P_jS_k
    =
    (-1)^{A_{jk}}S_kP_j,
    \qquad
    A_{jk}
    =
    \begin{cases}
        0, & [P_j,S_k]=0,\\
        1, & \{P_j,S_k\}=0.
    \end{cases}
\end{equation}
Commuting $P_j$ through $g(\bm q)$ gives
\begin{equation}
    P_jg(\bm q)
    =
    (-1)^{\sum_kA_{jk}q_k}
    g(\bm q)P_j
    =
    B_j(\bm\sigma)g(\bm q)P_j,
\end{equation}
where the sum in the exponent is evaluated modulo two and
\begin{equation}
    B_j(\bm\sigma)
    =
    \prod_{k=1}^{N_s}\sigma_k^{A_{jk}}.\label{5}
\end{equation}
Thus, each physical Pauli operator $P_j$ determines an
interaction $B_j$ in the classical spin model.  Its spatial
support consists precisely of the stabilizer generators that
anticommute with $P_j$.

The group multiplication also has a simple classical interpretation.
For two stabilizer configurations~$\boldsymbol\sigma$ and
$\boldsymbol s$, multiplication of the corresponding group elements
is represented by the componentwise product
\begin{equation}
    (\boldsymbol\sigma\boldsymbol s)_k
    :=
    \sigma_ks_k.
\end{equation}
Equation~\eqref{5} then implies
\begin{equation}
    B_j(\boldsymbol\sigma\boldsymbol s)
    =
    B_j(\boldsymbol\sigma)B_j(\boldsymbol s).
    \label{eq:multiplicative-property}
\end{equation}
This property will be the central algebraic ingredient in the
higher Nishimori identities derived in Sec.~\ref{app:generalized-HN}.

\label{subsec:stabilizer-relations}

The local stabilizer generators need not be independent.  Their
relations form the binary vector space
\begin{equation}
    \mathcal R
    =
    \left\{
        \boldsymbol r\in\mathbb F_2^{N_s}
        \;\middle|\;
        \prod_{k=1}^{N_s}S_k^{r_k}
        =
        \mathbb I
    \right\}.
    \label{eq:stabilizer-relation-space}
\end{equation}
If $\boldsymbol r\in\mathcal R$, the two binary configurations
$\boldsymbol q$ and $\boldsymbol q+\boldsymbol r$ describe the same
stabilizer element.  In terms of the Ising spins, this relation acts as
\begin{equation}
    \sigma_k
    \longmapsto
    (-1)^{r_k}\sigma_k.
    \label{eq:relation-spin-transformation}
\end{equation}
Since the operator in~\eqref{eq:stabilizer-relation-space} is the
identity, it commutes with every $P_j$.  It follows that
\begin{equation*}
    \sum_kA_{jk}r_k=0
    \pmod 2
\end{equation*}
and hence
\begin{equation}
    B_j\bigl(
        \{(-1)^{r_k}\sigma_k\}
    \bigr)
    =
    B_j(\boldsymbol\sigma).
    \label{eq:relation-invariance}
\end{equation}
Stabilizer relations therefore appear as exact symmetries of the
classical spin Hamiltonian.

It is convenient to retain all local stabilizer generators rather
than choosing a generally nonlocal independent subset.  The resulting
spin description contains a constant redundancy:
\begin{equation}
    \sum_{g\in\mathcal S}F(g)
    =
    \frac{1}{|\mathcal R|}
    \sum_{\boldsymbol\sigma\in\{\pm1\}^{N_s}}
    F\bigl(g(\boldsymbol\sigma)\bigr)
    \label{eq:lifted-spin-sum}
\end{equation}
for any function $F$ of the stabilizer element.  The factor
$|\mathcal R|$ only changes the overall normalization.  Physical
observables obtained from the stabilizer code must be invariant under~\eqref{eq:relation-spin-transformation}.

The relation between stabilizer identities and classical symmetries
will be important in the examples in later sections. 
The global relation among
toric code star operators produces the global Ising symmetry.  The
relations of the rank-2 toric code produce modulated symmetries (Sec.~\ref{222}),
while the subextensive relations of the X-cube code produce planar
subsystem symmetries (Appendix~\ref{xcube}).

\subsection{Classical Boltzmann weights from local Pauli noise}
\label{subsec:classical-weight}
We next consider local Pauli noise.
The classical Boltzmann weight can be seen directly from a local
Pauli channel. For a state $\rho_0$, we consider a channel induced by local Pauli noise with probability $p_j$: 
\begin{equation}
    \mathcal E_j[\rho_0 ]
    =
    (1-p_j)\rho_0+p_jP_j\rho_0 P_j,
    \qquad
    0\leq p_j<\frac12.
    \label{eq:local-Pauli-channel}
\end{equation}
Its action on a stabilizer element is given by
\begin{align}
    \mathcal E_j[g(\boldsymbol q)]
    &=
    \left[
        (1-p_j)+p_jB_j(\boldsymbol\sigma)
    \right]g(\boldsymbol q)
   =
    (1-2p_j)^{
        [1-B_j(\boldsymbol\sigma)]/2
    }
    g(\boldsymbol q).
    \label{eq:channel-on-stabilizer}
\end{align}
For independent noise channels, we introduce
\begin{equation}
    \mathcal E
    :=
    \prod_j\mathcal E_j.
\end{equation}
Defining
\begin{equation}
    \mu_j\vcentcolon=-\log(1-2p_j),
    \qquad
    K_j\vcentcolon=\frac{\mu_j}{2},
\end{equation}
the coefficient produced by independent local channels becomes
\begin{align}
    w(\boldsymbol\sigma)
    &=
    \prod_j
    \exp\left[
        -\frac{\mu_j}{2}
        \bigl(1-B_j(\boldsymbol\sigma)\bigr)
    \right]
   \propto
    \exp\left[
        \sum_jK_jB_j(\boldsymbol\sigma)
    \right].
    \label{eq:Boltzmann-weight}
\end{align}
The associated classical Hamiltonian is therefore
\begin{equation}
   H_{\mathrm{cl}}[\boldsymbol\sigma]
    =
    -\sum_jK_jB_j(\boldsymbol\sigma).
   \label{eq:general-classical-Hamiltonian}
\end{equation}
%
This construction depends only on the commutation algebra
between the local Pauli operators and the stabilizer generators.
In the following subsection, we show that the same interaction
functions $B_j(\bm\sigma)$ arise from the Born probabilities of
a deformed code state.  Weak measurements then promote the clean Boltzmann distribution to a posterior distribution conditioned on the measurement record.


\subsection{Wavefunction deformation and weak measurement}~\label{sc3}
We now show how the classical model introduced above arises from
a deformed stabilizer code wave function subjected to weak
measurements.  
%
We focus on Gaussian weak measurements,
for which the replicated statistical mechanical representation
derived below is exact at all length scales
\cite{patil2026higher}.

Since the operators $\{P_j\}$ mutually commute, we choose a
common reference state $|\Omega\rangle$ satisfying
\begin{equation}
    P_j|\Omega\rangle=|\Omega\rangle
    \qquad
    \text{for all }j.
\end{equation}
Acting with stabilizer group elements generates the orbit states
\begin{equation}
    |\bm\sigma\rangle
    :=
    \prod_k
    S_k^{(1-\sigma_k)/2}
    |\Omega\rangle,
    \qquad
    \sigma_k=\pm1.
\end{equation}
Configurations related by elements of $\mathcal R$ represent
the same orbit state and are therefore identified.  We assume
that the resulting inequivalent orbit states form an
orthonormal set.
In this
basis, we have
\begin{equation}
P_j\ket{\boldsymbol{\sigma}}
=
B_j(\boldsymbol{\sigma})
\ket{\boldsymbol{\sigma}},
\qquad
B_j(\boldsymbol{\sigma})
=
\prod_k \sigma_k^{A_{jk}}.
\label{eq:classical-monomial}
\end{equation}
Let $\mathcal C$ denote the set of inequivalent orbit
configurations.  We first define the uniform orbit state
\begin{equation}
    |\Psi_0\rangle
    =
    \frac{1}{\sqrt{|\mathcal C|}}
    \sum_{\bm\sigma\in\mathcal C}
    |\bm\sigma\rangle.
\end{equation}
We also introduce the classical partition function
\begin{equation}
    Z_{\bm K}
    :=
    \sum_{\bm\sigma\in\mathcal C}
    \exp\left[
        \sum_jK_jB_j(\bm\sigma)
    \right].
\end{equation}
Here, $K_j$ is a dimensionless coupling inherited from the wave function deformation and plays the role of an inverse temperature weighted coupling in the associated clean classical model. 
The normalized deformed code state is then
\begin{align}
    |\Psi_{\bm K}\rangle
    &:=
    \sqrt{\frac{|\mathcal C|}{Z_{\bm K}}}
    \exp\left[
        \frac12\sum_jK_jP_j
    \right]
    |\Psi_0\rangle
 =
    \frac{1}{\sqrt{Z_{\bm K}}}
    \sum_{\bm\sigma\in\mathcal C}
    \exp\left[
        \frac12\sum_jK_jB_j(\bm\sigma)
    \right]
    |\bm\sigma\rangle.
\end{align}

A projective measurement in the orbit basis therefore gives
\begin{equation}
    P_{\bm K}(\bm\sigma)
    =
    |\langle\bm\sigma|\Psi_{\bm K}\rangle|^2
    =
    \frac{1}{Z_{\bm K}}
    \exp\left[
        \sum_jK_jB_j(\bm\sigma)
    \right].
\end{equation}
Thus, the Born probabilities of the deformed code state coincide
with the Boltzmann distribution of the associated clean
classical model.

We now subject the operators $P_j$ to independent Gaussian weak
measurements.  For a measurement record
$\bm m=\{m_j\}$, the joint Kraus operator is described by
\begin{equation}
    M_{\bm m}
    =
    \prod_j
    \left(
        \frac{\Delta_j}{2\pi}
    \right)^{1/4}
    \exp\left[
        -\frac{\Delta_j}{4}
        (m_j-P_j)^2
    \right],
    \qquad
    m_j\in\mathbb R.
\end{equation}
The Kraus operators satisfy
\begin{equation}
    \int\mathrm d\bm m\,
    M_{\bm m}^{\dagger}M_{\bm m}
    =
    \mathbb I,
    \qquad
    \mathrm d\bm m:=\prod_j\mathrm dm_j.
\end{equation}
Conditioned on an orbit configuration $\bm\sigma$, the
measurement outcomes are independently distributed according to
\begin{equation}
    P_{\bm\Delta}(\bm m|\bm\sigma)
    =
    \prod_j
    \left(
        \frac{\Delta_j}{2\pi}
    \right)^{1/2}
    \exp\left[
        -\frac{\Delta_j}{2}
        \left(
            m_j-B_j(\bm\sigma)
        \right)^2
    \right].
\end{equation}
Bayes' theorem gives the posterior distribution
\begin{equation}
    P(\bm\sigma|\bm m)
    =
    \frac{1}{Z[\bm m]}
    \exp\left[
        \sum_j
        \left(
            K_j+\Delta_jm_j
        \right)
        B_j(\bm\sigma)
    \right],
    \label{eq:posterior-distribution}
\end{equation}
where the record dependent partition function is
\begin{equation}
    Z[\bm m]
    :=
    \sum_{\bm\sigma\in\mathcal C}
    \exp\left[
        \sum_j
        \left(
            K_j+\Delta_jm_j
        \right)
        B_j(\bm\sigma)
    \right].
    \label{eq:record-partition-function}
\end{equation}
In the chosen stabilizer orbit basis, the normalized
post measurement state is therefore
\begin{equation}
    |\Psi_{\bm m}\rangle
    =
    \sum_{\bm\sigma\in\mathcal C}
    \sqrt{
        P(\bm\sigma|\bm m)
    }
    |\bm\sigma\rangle.
    \label{eq:postmeasurement-state}
\end{equation}
Consequently, posterior averages of $B_j(\bm\sigma)$ coincide
with quantum expectation values of the corresponding diagonal
operators $P_j$ in $|\Psi_{\bm m}\rangle$.

The Born probability density of the measurement record is
\begin{equation}
    P(\bm m)
    =
    \frac{g_{\bm\Delta}(\bm m)}{Z_{\bm K}}
    Z[\bm m],
    \label{eq:record-probability}
\end{equation}
where
\begin{equation}
    g_{\bm\Delta}(\bm m)
    :=
    \prod_j
    \left(
        \frac{\Delta_j}{2\pi}
    \right)^{1/2}
    \exp\left[
        -\frac{\Delta_j}{2}
        \left(
            m_j^2+1
        \right)
    \right].
    \label{eq:gaussian-record-weight}
\end{equation}
Note that record averaged products of posterior expectation values contain
inverse powers of the normalization factor $Z[\bm m]$.  
For an observable $O(\bm\sigma)$, we denote its posterior expectation
at a fixed measurement record by
\begin{eqnarray*}
     \langle O\rangle_{\bm m}
 :=
 \sum_{\bm\sigma\in\mathcal C}
 P(\bm\sigma|\bm m)\,O(\bm\sigma).
\end{eqnarray*}
The average over measurement records with their
Born probabilities is denoted by an overline, namely, 
\begin{eqnarray}
     \overline{F(\bm m)}
 :=
 \int \mathrm d\bm m\,P(\bm m)\,F(\bm m).
\end{eqnarray}
Introducing $R>0$ replicas of the
posterior statistical model, we define
the replicated generating function as
\begin{equation}
    \mathcal Z_R
    :=
    \frac{1}{Z_{\bm K}}
    \int \mathrm d\bm m\,
    g_{\bm\Delta}(\bm m)
    Z[\bm m]^R,
    \qquad
    \mathrm d\bm m:=\prod_j\mathrm dm_j.
    \label{eq:replicated-generating-function}
\end{equation}
The physical Born rule average is obtained by analytically
continuing $R$ to one.  This differs from the conventional
quenched-disorder replica limit $R\to0$ because the Born
probability $P(\bm m)$ itself contains one power of
$Z[\bm m]$, as shown in~\eqref{eq:record-probability}.  In particular,
\begin{equation}
    \mathcal Z_1
    =
    \int\mathrm d\bm m\,P(\bm m)
    =
    1.
\end{equation}
For integer $R$, the replicated power can be written as
\begin{align}
    Z[\bm m]^R
    =
    \sum_{\{
        \bm\sigma^{(a)}
    \}_{a=1}^{R}}
    \exp\Bigg[
        \sum_j\sum_{a=1}^{R}
        \left(
            K_j+\Delta_jm_j
        \right)
        B_j\bigl(\bm\sigma^{(a)}\bigr)
    \Bigg].
    \label{eq:replicated-record-partition-function}
\end{align}
Here $a=1,\ldots,R$ is the replica index.  For each measurement
outcome $m_j$, the Gaussian integral is
\begin{align}
&\int_{-\infty}^{\infty}\mathrm dm_j
\left(
    \frac{\Delta_j}{2\pi}
\right)^{1/2}
\exp\left[
    -\frac{\Delta_j}{2}
    \left(
        m_j^2+1
    \right)
    +
    \Delta_jm_j
    \sum_{a=1}^{R}
    B_j^{(a)}
\right]
=
\exp\left[
    -\frac{\Delta_j}{2}
    +
    \frac{\Delta_j}{2}
    \left(
        \sum_{a=1}^{R}B_j^{(a)}
    \right)^2
\right],
\label{eq:gaussian-replica-integral}
\end{align}
where
 $ B_j^{(a)}
    :=
    B_j\bigl(\bm\sigma^{(a)}\bigr).$
Substituting~\eqref{eq:gaussian-replica-integral} into~\eqref{eq:replicated-generating-function}, we obtain
\begin{align}
    \mathcal Z_R
    =
    \frac{
        e^{-\frac12\sum_j\Delta_j}
    }{
        Z_{\bm K}
    }
    \sum_{\{
        \bm\sigma^{(a)}
    \}_{a=1}^{R}}
    \exp\Bigg[
        &
        \sum_j\sum_{a=1}^{R}
        K_jB_j^{(a)}
     +
        \frac12\sum_j\Delta_j
        \left(
            \sum_{a=1}^{R}B_j^{(a)}
        \right)^2
    \Bigg].
    \label{eq:replicated-partition-function}
\end{align}
Since $\bigl(B_j^{(a)}\bigr)^2=1$, we have
\begin{equation}
    \left(
        \sum_{a=1}^{R}B_j^{(a)}
    \right)^2
    =
    R+
    \sum_{\substack{a,b=1\\a\neq b}}^{R}
    B_j^{(a)}B_j^{(b)}.
\end{equation}
Therefore, after dropping terms independent of the spin
configurations, the replicated Hamiltonian is given by
\begin{equation}
    -H_R
    =
    \sum_j\sum_{a=1}^{R}
    K_jB_j^{(a)}
    +
    \frac12
    \sum_j\Delta_j
    \sum_{\substack{a,b=1\\a\neq b}}^{R}
    B_j^{(a)}B_j^{(b)}.
    \label{eq:replicated-Hamiltonian}
\end{equation}
We now focus on the parameter manifold
\begin{equation}
    \Delta_j=K_j\;\forall j
    \label{eq:higher-Nishimori-manifold}
\end{equation}
which we call the \emph{higher Nishimori manifold}.  This condition matches the coupling $K_j$ inherited from the
wavefunction deformation with the coupling $\Delta_j$ between
replicas generated by the measurement.  It is the analog, in the present learning
problem, of the matching between the prior and likelihood on
the conventional Nishimori line.

The special role of~\eqref{eq:higher-Nishimori-manifold} can already be seen
from the replicated Hamiltonian.  On this manifold, it can be
written, up to a spin-independent constant, as
\begin{equation}
    -H_R
    =
    \frac12
    \sum_jK_j
    \left[
        1+
        \sum_{a=1}^{R}
        B_j\bigl(\bm\sigma^{(a)}\bigr)
    \right]^2
    +\mathrm{const}.
    \label{eq:HN-perfect-square}
\end{equation}
Away from the higher Nishimori manifold, the single replica and
inter replica terms have independent coefficients and cannot be
combined into this form. There is an additional replica symmetry in~\eqref{eq:HN-perfect-square}.  To see this, we introduce an auxiliary Ising
configuration $\bm s=\{s_k\}$, average uniformly over
$\bm s$, and perform the change of variables
\begin{equation}
    \bm\sigma^{(a)}
    \longmapsto
    \bm\sigma^{(a)}\bm s,
    \qquad
    a=1,\ldots,R.
\end{equation}
Using the multiplicative property
\begin{equation}
    B_j(\bm\sigma\bm s)
    =
    B_j(\bm\sigma)B_j(\bm s),
\end{equation}
we find
\begin{align}
    \left[
        1+
        \sum_{a=1}^{R}
        B_j\bigl(\bm\sigma^{(a)}\bm s\bigr)
    \right]^2
    &=
    \left[
        B_j(\bm s)+
        \sum_{a=1}^{R}
        B_j\bigl(\bm\sigma^{(a)}\bigr)
    \right]^2.
\end{align}
Identifying
   $ \bm\sigma^{(R+1)}
    :=
    \bm s,$
the replicated partition function can therefore be rewritten,
up to an overall normalization, as
\begin{equation}
    \mathcal Z_R
    \propto
    \sum_{\{
        \bm\sigma^{(a)}
    \}_{a=1}^{R+1}}
    \exp\left[
        \frac12
        \sum_jK_j
        \left(
            \sum_{a=1}^{R+1}
            B_j\bigl(\bm\sigma^{(a)}\bigr)
        \right)^2
    \right].
    \label{eq:HN-R-plus-one}
\end{equation}
This representation has a manifest enlarged replica
permutation symmetry
   $ S_R
    \longrightarrow
    S_{R+1},$
which treats the auxiliary configuration, interpreted as the
truth replica, on the same footing as the $R$ posterior
replicas. In the next section, we make use of these properties to derive several useful identities.  
\section{Higher Nishimori identities}
\label{app:generalized-HN}
We now use the symmetry derived in Sec.~\ref{sec:classical-mapping} to relate the learnability of different observables. In particular, we ask how learning two observables constrains the learnability of their product. We first review the ordinary Nishimori condition~\cite{Nishimori_1980,iba1999nishimori}. This will clarify the role of the additional replica before we turn to the weak measurement problem.
We then derive an identity involving several observables and their product. Finally, we combine this identity with the Griffiths–Kelly–Sherman (GKS) correlation inequalities~\cite{Griffiths_1967,Kelly_Sherman_1968} in the clean model to obtain a hierarchy of learnability.

\subsection{Ordinary Nishimori condition}
We first review the ordinary Nishimori condition for the classical spin models introduced in Sec.~\ref{sec:classical-mapping}~\footnote{We thank Bo Han for discussing  this issue.}. 
We consider quenched binary disorder $\tau_j=\pm1$ coupled to the local
interactions $B_j(\sigma)$,
\begin{equation}
    H_{\tau}(\sigma)
    =
    -\sum_j K_j \tau_j B_j(\sigma),
\end{equation}
with independent disorder distribution
\begin{equation}
    P_{\kappa}(\tau)
    =
    \prod_j
    \frac{e^{\kappa_j \tau_j}}
         {2\cosh \kappa_j}. \label{eqn:disorderP}
\end{equation}
The signs are held fixed when taking a thermal expectation value and averaged over afterwards. We denote these two averages by $\langle\cdot\rangle_{\boldsymbol{\tau}}$ and $[\cdot]_{\boldsymbol{\tau}}$, respectively. If $p_j=P_{\kappa}(\tau_j=-1)$, then~\eqref{eqn:disorderP} gives $\kappa_j=\frac12\log[(1-p_j)/p_j]$. The ordinary Nishimori condition sets this disorder parameter equal to the thermal coupling:
\begin{equation}
    \kappa_j = K_j
    \qquad \forall j.\label{48}
\end{equation}
Averaging the Boltzmann weight of $n$ replicas over $\tau_j$ gives the factor $\cosh[\kappa_j+K_j\sum_{a=1}^{n}B_j(\boldsymbol{\sigma}^{(a)})]/\cosh\kappa_j$. Under the condition~\eqref{48}, the disorder bias and each replica contribute with the same coefficient. The replicated partition function therefore takes the following form:
\begin{equation}
    [Z_{\tau}^{\,n}]_{\tau}
    \propto
    \sum_{\{\sigma^{(a)}\}}
    \prod_j
    \cosh\left[
        K_j
        \left(
            1+\sum_{a=1}^{n}B_j(\sigma^{(a)})
        \right)
    \right].
\end{equation}
Introducing an arbitrary spin configuration $s$ and using the
evenness of the hyperbolic cosine yields
\begin{equation}
    \cosh\left[
        K_j
        \left(
            1+\sum_{a=1}^{n}B_j(\sigma^{(a)}s)
        \right)
    \right]
    =
    \cosh\left[
        K_j
        \left(
            B_j(s)+\sum_{a=1}^{n}B_j(\sigma^{(a)})
        \right)
    \right].
\end{equation}
Averaging uniformly over $\boldsymbol s$ and identifying it with $\boldsymbol{\sigma}^{(n+1)}$ gives the following replicated theory up to an overall normalization:
\begin{equation}
    [Z_{\tau}^{\,n}]_{\tau}
    \propto
    \sum_{\{\sigma^{(a)}\}_{a=1}^{n+1}}
    \prod_j
    \cosh\left[
        K_j
        \sum_{a=1}^{n+1}B_j(\sigma^{(a)})
    \right],
\end{equation}
which exhibits the familiar enlarged permutation symmetry
$S_n\rightarrow S_{n+1}$ together with the local gauge redundancy
\begin{equation}
    \sigma_k^{(a)}
    \longrightarrow
    \eta_k \sigma_k^{(a)},
    \qquad
    a=1,\ldots,n+1
\end{equation}
with $\eta_k=\pm1$.
For a binary vector $\boldsymbol v\in\mathbb F_2^{N_s}$, we define the spin product
\begin{eqnarray}
    O_{\boldsymbol v}(\boldsymbol\sigma):=\prod_k\sigma_k^{v_k}.\label{ov}
\end{eqnarray}
These observables satisfy $O_{\boldsymbol v}(\boldsymbol\sigma\boldsymbol s)=O_{\boldsymbol v}(\boldsymbol\sigma)O_{\boldsymbol v}(\boldsymbol s)$ and $O_{\boldsymbol v}O_{\boldsymbol w}=O_{\boldsymbol v\oplus\boldsymbol w}$, where $\oplus$ denotes componentwise addition modulo two. Combining these properties with replica permutations and local gauge transformations gives the following identity:
\begin{equation}
    \left[
    \prod_{\lambda=1}^{M}
    \langle O_{v_\lambda}\rangle_{\tau}
    \right]_{\tau}
    =
    \left[
    \left(
    \prod_{\lambda=1}^{M}
    \langle O_{v_\lambda}\rangle_{\tau}
    \right)
    \langle O_q\rangle_{\tau}
    \right]_{\tau},
    \qquad
    q=\bigoplus_{\lambda=1}^{M}v_\lambda.
    \label{eq:ordinary_heterogeneous_Nishimori}
\end{equation}
For an odd number of identical insertions,
$v_1=\cdots=v_{2m-1}=v$, this reduces to the identity
\begin{equation}
    \left[
    \langle O_v\rangle_{\tau}^{2m-1}
    \right]_{\tau}
    =
    \left[
    \langle O_v\rangle_{\tau}^{2m}
    \right]_{\tau}.
\end{equation}
The learning problem has a similar replica symmetry, but a different physical interpretation. In the ordinary problem, the additional replica arises from the disorder distribution, and averages over quenched disorder are obtained in the limit $n\to0$. In the learning problem, the additional replica represents the unknown spin configuration, and averages over measurement records are obtained in the limit $R\to1$. We now derive the corresponding identity for learning. Its connection to correlations in the clean model will be established in Sec.~\ref{subsec:exact-correlator-hierarchy}.
\subsection{Relation among record averaged correlators}
We now return to the learning problem on the higher Nishimori manifold, $\Delta_j=K_j$. We restrict to observables that are invariant under the stabilizer relations. Let $O_{\boldsymbol v_1},\ldots,O_{\boldsymbol v_M}$ be $M$ such observables, which need not be distinct. We place them in different replicas with indices $a_1,\ldots,a_M$, and define the corresponding replica correlator:
\begin{equation*}
    {\cal C}_R(\bm v_1,\ldots,\bm v_M)
    :=
    \left\langle
        \prod_{\lambda=1}^{M}
        O_{\bm v_\lambda}
        (\bm\sigma^{(a_\lambda)})
    \right\rangle_R .
 \end{equation*}
Here $\langle\cdot\rangle_R$ denotes the normalized expectation value in the replicated model.
Explicitly,~\footnote{Throughout this section, each replica is summed over all
$2^{N_s}$ spin configurations, including the redundancy
associated with $\mathcal R$.
The spin partition function $Z_R$ is distinct from the
generating function $\mathcal Z_R$. On the higher Nishimori manifold,
they are related by
$\mathcal Z_R
    =  \frac{e^{-\sum_jK_j}}{Z_K|\mathcal R|^R}\,Z_R.$ The extra factors are independent of the spin configurations
and cancel in normalized replica expectation values.}
\begin{align}
    {\cal C}_R
    &=
    \frac{1}{Z_R}
    \sum_{\{\bm\sigma^{(a)}\}}
    \prod_{\lambda=1}^{M}
    O_{\bm v_\lambda}
    (\bm\sigma^{(a_\lambda)})
    \times
    \exp\left[
        \frac{1}{2}
        \sum_j\Delta_j
        \left(
            1+
            \sum_{a=1}^{R}
            B_j(\bm\sigma^{(a)})
        \right)^2
    \right].
    \label{app:eq:C-explicit}
\end{align}
To proceed, we  introduce an arbitrary Ising configuration
$\bm s=\{s_k\}$ and average over all its configurations. Since
the original expression is independent of $\bm s$, we may write
\begin{align}
    {\cal C}_R
    &=
    \frac{1}{2^{N_s}Z_R}
    \sum_{\bm s}
    \sum_{\{\bm\sigma^{(a)}\}}
    \prod_{\lambda=1}^{M}
    O_{\bm v_\lambda}
    (\bm\sigma^{(a_\lambda)})
   \times
    \exp\left[
        \frac{1}{2}
        \sum_j\Delta_j
        \left(
            1+
            \sum_{a=1}^{R}
            B_j(\bm\sigma^{(a)})
        \right)^2
    \right].
    \label{app:eq:add-s}
\end{align}
For each fixed $\boldsymbol s$, make the invertible change of variables $\widetilde{\sigma}^{(a)}_k=s_k\sigma^{(a)}_k$. The multiplication rules for the interactions and observables then give
\begin{align}
    B_j(\widetilde{\bm\sigma}^{(a)})
   =
    B_j(\bm s)
    B_j(\bm\sigma^{(a)}),
\quad
    O_{\bm v_\lambda}
    (\widetilde{\bm\sigma}^{(a_\lambda)})
=
    O_{\bm v_\lambda}(\bm s)
    O_{\bm v_\lambda}
    (\bm\sigma^{(a_\lambda)}).
\end{align}
The product of the observable factors generated by the change of
variables is
\begin{align}
    \prod_{\lambda=1}^{M}
    O_{\bm v_\lambda}(\bm s)
    &=
    O_{\bm q}(\bm s),
    \label{app:eq:q-factor}
\end{align}
where
   $ \bm q
    :=
    \bm v_1\oplus
    \bm v_2\oplus\cdots\oplus
    \bm v_M .$
Moreover, since $B_j(\bm s)^2=1$, the Boltzmann weight transforms
according to
\begin{align}
    &
    \left[
        1+
        \sum_{a=1}^{R}
        B_j(\widetilde{\bm\sigma}^{(a)})
    \right]^2
    =
    \left[
        1+
        B_j(\bm s)
        \sum_{a=1}^{R}
        B_j(\bm\sigma^{(a)})
    \right]^2
    =
    \left[
        B_j(\bm s)
        +
        \sum_{a=1}^{R}
        B_j(\bm\sigma^{(a)})
    \right]^2 .
    \label{app:eq:weight-transform}
\end{align}

Dropping the tilde on the dummy summation variables, we therefore
find
\begin{align}
    {\cal C}_R
    &=
    \frac{1}{2^{N_s}Z_R}
    \sum_{\bm s}
    \sum_{\{\bm\sigma^{(a)}\}}
    \left[
        \prod_{\lambda=1}^{M}
        O_{\bm v_\lambda}
        (\bm\sigma^{(a_\lambda)})
    \right]
    O_{\bm q}(\bm s)
   \times
    \exp\left[
        \frac{1}{2}
        \sum_j\Delta_j
        \left(
            B_j(\bm s)+
            \sum_{a=1}^{R}
            B_j(\bm\sigma^{(a)})
        \right)^2
    \right].
    \label{app:eq:before-extra-replica}
\end{align}

We now identify the auxiliary spin configuration with an
additional replica,
  $ \bm\sigma^{(R+1)}:=\bm s.$
The correlator~\eqref{app:eq:before-extra-replica} becomes
\begin{align}
    {\cal C}_R
    &=
    \frac{1}{2^{N_s}Z_R}
    \sum_{\{\bm\sigma^{(a)}\}_{a=1}^{R+1}}
    \left[
        \prod_{\lambda=1}^{M}
        O_{\bm v_\lambda}
        (\bm\sigma^{(a_\lambda)})
    \right]
    O_{\bm q}(\bm\sigma^{(R+1)})
  \times
    \exp\left[
        \frac{1}{2}
        \sum_j\Delta_j
        \left(
            \sum_{a=1}^{R+1}
            B_j(\bm\sigma^{(a)})
        \right)^2
    \right].
    \label{app:eq:gauge-invariant-correlator}
\end{align}
The Boltzmann weight in~\eqref{app:eq:gauge-invariant-correlator} is manifestly
invariant under the enlarged replica permutation group~$ S_{R+1}$.
It is also invariant under the local transformation
\begin{equation}
    \sigma_k^{(a)}
    \longrightarrow
    \eta_k\sigma_k^{(a)},
    \qquad
    a=1,\ldots,R+1,
    \qquad
    \eta_k=\pm1 .
    \label{app:eq:local-gauge}
\end{equation}
Indeed, under such a transformation, we have
\begin{equation}
    B_j(\bm\sigma^{(a)})
    \longrightarrow
    \chi_j(\bm\eta)
    B_j(\bm\sigma^{(a)}),
\end{equation}
where
\begin{equation}
    \chi_j(\bm\eta)
    :=
    \prod_k\eta_k^{A_{jk}}
    =\pm1 ,
\end{equation}
so that
   $ \left[
        \sum_{a=1}^{R+1}
        B_j(\bm\sigma^{(a)})
    \right]^2$
is unchanged.

Notice also that the complete operator insertion in~\eqref{app:eq:gauge-invariant-correlator} is gauge invariant.
Under the change~\eqref{app:eq:local-gauge}, it acquires the factor
\begin{equation}
    \prod_{\lambda=1}^{M}
    O_{\bm v_\lambda}(\bm\eta)
    O_{\bm q}(\bm\eta)
    =
    O_{\bm q}(\bm\eta)^2
    =
    1.
\end{equation}

We choose a replica index $b$ satisfying
 $  b\neq a_1,\ldots,a_M,
    \;
    1\leq b\leq R.$
For integer $R$ sufficiently large such a replica always exists.
Using the $S_{R+1}$ symmetry, we exchange replicas $b$ and
$R+1$. Since replica~$b$ initially carries no operator insertion,~\eqref{app:eq:gauge-invariant-correlator} becomes
\begin{align}
    {\cal C}_R
    &=
    \frac{1}{2^{N_s}Z_R}
    \sum_{\{\bm\sigma^{(a)}\}_{a=1}^{R+1}}
    \left[
        \prod_{\lambda=1}^{M}
        O_{\bm v_\lambda}
        (\bm\sigma^{(a_\lambda)})
    \right]
    O_{\bm q}(\bm\sigma^{(b)})
    \times
    \exp\left[
        \frac{1}{2}
        \sum_j\Delta_j
        \left(
            \sum_{a=1}^{R+1}
            B_j(\bm\sigma^{(a)})
        \right)^2
    \right].
    \label{app:eq:after-permutation}
\end{align}
We finally fix the local gauge by imposing
 $\sigma_k^{(R+1)}=+1\;
    \forall k.$ Since
    $B_j(\bm 1)=1$,
the Boltzmann weight reduces precisely to the original
$R$-replica weight. The sum over the
gauge orbit gives a factor~$2^{N_s}$, which cancels the explicit
factor~$2^{-N_s}$ in~\eqref{app:eq:after-permutation}. We therefore obtain
\begin{equation}
   \left\langle
        \prod_{\lambda=1}^{M}
        O_{\bm v_\lambda}^{(a_\lambda)}
    \right\rangle_R
    =
    \left\langle
        \left[
        \prod_{\lambda=1}^{M}
        O_{\bm v_\lambda}^{(a_\lambda)}
        \right]
        O_{\bm q}^{(b)}
    \right\rangle_R ,
    \label{app:eq:heterogeneous-replica-identity}
\end{equation}
where
$     \bm q=
    \bigoplus_{\lambda=1}^{M}\bm v_\lambda$ and $O_{\boldsymbol v}^{(a)}$ is shorthand for $O_{\boldsymbol v}(\boldsymbol\sigma^{(a)})$.
The identity applies to distinct observables as well as repeated copies of the same observable. 
Products of posterior expectation values are represented by
insertions into pairwise distinct replicas. In particular,
\begin{equation}
    \overline{
        \prod_{\lambda=1}^{M}
        \langle O_{\bm v_\lambda}\rangle_{\bm m}
    }
    =
    \lim_{R\rightarrow1}
    \left\langle
        \prod_{\lambda=1}^{M}
        O_{\bm v_\lambda}^{(a_\lambda)}
    \right\rangle_R .
    \label{app:eq:learning-replica}
\end{equation}
Analytically continuing
the relation~\eqref{app:eq:heterogeneous-replica-identity} to the
learning limit therefore yields
\begin{equation}
    \boxed{
    \overline{
        \prod_{\lambda=1}^{M}
        \langle O_{\bm v_\lambda}\rangle_{\bm m}
    }
    =
    \overline{
        \left[
        \prod_{\lambda=1}^{M}
        \langle O_{\bm v_\lambda}\rangle_{\bm m}
        \right]
        \langle O_{\bm q}\rangle_{\bm m}
    },
    \qquad
    \bm q=
    \bigoplus_{\lambda=1}^{M}\bm v_\lambda .
    }
    \label{app:eq:heterogeneous-learning}
\end{equation}
For the Gaussian measurements considered here, this is an exact identity. It can also be obtained directly from the joint distribution of the spin configuration and measurement record, without continuing in the replica number. For binary measurements, the corresponding relation is expected at long distances if the model approaches the same higher Nishimori fixed point.

As a consistency check, consider an odd number $M=2k-1$ of
identical observables,
\begin{equation}
    \bm v_1=\cdots=\bm v_{2k-1}=\bm v.
\end{equation}
Since addition is modulo two,
\begin{equation}
    \bm q
    =
    \underbrace{
    \bm v\oplus\cdots\oplus\bm v
    }_{2k-1\ {\rm times}}
    =
    \bm v .
\end{equation}
The relation~\eqref{app:eq:heterogeneous-learning} then reduces to $\overline{
        \langle O_{\bm v}\rangle_{\bm m}^{2k-1}
    }
    =
    \overline{
        \langle O_{\bm v}\rangle_{\bm m}^{2k}
    } $, which reproduces the higher Nishimori identity~\cite{patil2026higher}. 
\subsection{Exact hierarchy of record averaged correlators}
\label{subsec:exact-correlator-hierarchy}
We now use~\eqref{app:eq:heterogeneous-learning} to relate learnability to correlations in the clean model. For an observable $O_{\boldsymbol v}$, define the average of its squared posterior expectation by
\begin{equation}
    \mathcal C_{\bm v}^{(2)}
    :=
    \overline{
        \left\langle O_{\bm v}\right\rangle_{\bm m}^{2}
    }.
    \label{eq:second-moment-correlator}
\end{equation}
When $O_{\bm v}$ is a spatial correlation operator,
$\mathcal C_{\bm v}^{(2)}$ is its Edwards--Anderson (EA)-type correlator~\cite{S_Edwards_1975,Castellani_2005}
and measures how reliably the corresponding nonlocal information can
be inferred from the measurement record.

Taking $M=1$ in~\eqref{app:eq:heterogeneous-learning} gives
\begin{equation}
    \overline{
        \left\langle O_{\bm v}\right\rangle_{\bm m}
    }
    =
    \mathcal C_{\bm v}^{(2)}.
    \label{eq:first-second-HN}
\end{equation}
Averaging a posterior expectation over all possible records recovers the expectation in the original distribution:
\begin{align}
    \overline{
        \left\langle O_{\bm v}\right\rangle_{\bm m}
    }
    & =
    \int \mathrm{d}\bm m\,P(\bm m)
    \sum_{\bm\sigma}
    P(\bm\sigma|\bm m)O_{\bm v}(\bm\sigma)
   =
    \sum_{\bm\sigma}
    P_{\bm K}(\bm\sigma)O_{\bm v}(\bm\sigma)
    =:
    \left\langle O_{\bm v}\right\rangle_{\rm clean}.
    \label{eq:total-expectation-clean}
\end{align}
Combining~\eqref{eq:first-second-HN} and
\eqref{eq:total-expectation-clean}, we obtain the general equality
\begin{equation}
    \boxed{
    \mathcal C_{\bm v}^{(2)}
    =
    \left\langle O_{\bm v}\right\rangle_{\rm clean}.
    }
    \label{eq:second-moment-clean-general}
\end{equation}
Therefore, on the higher Nishimori manifold, the average squared posterior expectation is equal to the corresponding correlation in the clean model. Here the clean model is the distribution $P_K$ associated with the deformed state before conditioning on a measurement record.

We next assume that all couplings $K_j$ multiplying the spin products $B_j$ are nonnegative. In this sense, the clean model is ferromagnetic even when its interactions involve more than two spins. The GKS inequalities then give
\begin{equation}
    \left\langle
        O_{\bm v}O_{\bm w}
    \right\rangle_{\rm clean}
    \geq
    \left\langle O_{\bm v}\right\rangle_{\rm clean}
    \left\langle O_{\bm w}\right\rangle_{\rm clean}
    \label{eq:GKS-two-observables}
\end{equation}
for arbitrary product of spins $O_{\bm v}$ and $O_{\bm w}$~\cite{Griffiths_1967,Kelly_Sherman_1968} (See Appendix~\ref{gks} for a review. ).  
In words, the expectation of their product is at least the product of their expectations.
Since
$O_{\bm v}O_{\bm w}=O_{\bm v\oplus\bm w}$,~\eqref{eq:second-moment-clean-general}
converts the inequality in the clean model directly into a hierarchy among
learning correlators, leading to the following:
\begin{frm-thm}[Exact correlator hierarchy]\label{th}
On the higher Nishimori manifold of a CSS stabilizer code learning
problem whose associated clean spin model has nonnegative
ferromagnetic couplings,
\begin{equation}
    \boxed{
    \mathcal C_{\bm v\oplus\bm w}^{(2)}
    \geq
    \mathcal C_{\bm v}^{(2)}
    \mathcal C_{\bm w}^{(2)}.
    }
    \label{eq:exact-correlator-hierarchy}
\end{equation}
More generally, for any collection
$\bm v_1,\ldots,\bm v_M$,
\begin{equation}
    \boxed{
    \mathcal C_{\bigoplus_{\lambda=1}^{M}\bm v_\lambda}^{(2)}
    \geq
    \prod_{\lambda=1}^{M}
    \mathcal C_{\bm v_\lambda}^{(2)}.
    }
    \label{eq:exact-correlator-hierarchy-M}
\end{equation}
\end{frm-thm}

The inequality for two observables follows directly from \eqref{eq:second-moment-clean-general} and~\eqref{eq:GKS-two-observables}.  Repeating the same argument gives~\eqref{eq:exact-correlator-hierarchy-M}, since each intermediate product is again of the form~\eqref{ov}. These inequalities relate the learnability of different observables to that of their product.
The hierarchy immediately constrains long range order.  Suppose that
$O_{\bm v}(r)$ and $O_{\bm w}(r)$ are families of correlation operators at separation $r$, and define
their product channel by
$O_{\bm q}(r)=O_{\bm v}(r)O_{\bm w}(r)$, with
$\bm q=\bm v\oplus\bm w$.  If
\begin{equation}
    q_{\bm v}
    :=
    \lim_{r\rightarrow\infty}\mathcal C_{\bm v}^{(2)}(r)>0,
    \qquad
    q_{\bm w}
    :=
    \lim_{r\rightarrow\infty}\mathcal C_{\bm w}^{(2)}(r)>0,
\end{equation}
then
\begin{equation}
    q_{\bm q}
    \geq
    q_{\bm v}q_{\bm w}>0.
    \label{eq:order-hierarchy}
\end{equation}
Therefore, if both observables remain learnable at long distances, their product must also remain learnable. The inequality does not imply the converse: the product may remain learnable even if one or both of the original observables do not.
The same result gives a universal constraint at a continuous
transition.  Assume that the three second moment correlators have the
asymptotic power law behaviors
\begin{equation}
    \mathcal C_{\bm a}^{(2)}(r)
    \sim
    A_{\bm a}r^{-2x_{\bm a}^{(2)}},
    \qquad
    \bm a=\bm v,\bm w,\bm q,
    \qquad A_{\bm a}>0.
    \label{eq:critical-decay-general}
\end{equation}
The relation~\eqref{eq:exact-correlator-hierarchy} can hold at
arbitrarily large $r$ only if
\begin{equation}
    \boxed{
    x_{\bm v\oplus\bm w}^{(2)}
    \leq
    x_{\bm v}^{(2)}+x_{\bm w}^{(2)}.
    }
    \label{eq:scaling-dimension-hierarchy}
\end{equation}
This bound does not assume a particular universality class and remains valid even when the individual scaling dimensions are not known exactly.
\begin{table*}[t]
\centering
\footnotesize
\renewcommand{\arraystretch}{1.15}
\setlength{\tabcolsep}{4pt}

\caption{
Comparison between the ordinary and higher Nishimori structures.
}
\label{tab:nishimori_comparison}

\begin{tabular}{
p{0.15\textwidth}
p{0.36\textwidth}
p{0.40\textwidth}
}
\toprule
&
\textbf{Ordinary Nishimori}
&
\textbf{Higher Nishimori}
\\
\midrule

Setting
&
Quenched disorder / decoherence
&
Bayesian learning from weak measurements
\\

Replica limit
&
$n\to0$
&
$R\to1$
\\

Matching condition
&
Disorder distribution matched to thermal coupling
&
Measurement strength matched to prior coupling,
$\Delta_j=K_j$
\\

Symmetry
&
Gauge symmetry and
$S_n\to S_{n+1}$
&
Local gauge redundancy and
$S_R\to S_{R+1}$
\\

Consequence
&
$\displaystyle
[\langle O\rangle]_\tau
=
[\langle O\rangle^2]_\tau
$
&
$\displaystyle
C_{v\oplus w}^{(2)}
\geq
C_v^{(2)}C_w^{(2)}
\quad\text{(with GKS)}
$
\\

\bottomrule
\end{tabular}
\end{table*}
\par
In Table~\ref{tab:nishimori_comparison}, we summarize the comparison with the ordinary Nishimori problem. In both settings, replica symmetry and the multiplication rules of observables lead to identities among averaged correlators. In the learning problem,~\eqref{eq:second-moment-clean-general} further relates these quantities to correlations in the clean model. For nonnegative couplings, the GKS inequalities then give the hierarchy~\eqref{eq:exact-correlator-hierarchy-M}, which constrains the simultaneous learnability of distinct observables.

\section{Examples of CSS codes}\label{exx}
In this section, we  apply the higher Nishimori identities to concrete stabilizer code models. After reviewing learning problem of the toric code, we apply the general framework to the rank-2 toric code, whose associated classical model possesses modulated symmetries. 
\subsection{Warm-up: Toric code (review)}

We first consider the two dimensional toric code with qubits on
the links of a square lattice,
\begin{equation}
    H_{\rm TC}
    =
    -\sum_v A_v
    -\sum_p B_p,
\end{equation}
where
\begin{equation*}
    A_v=\prod_{e\ni v}X_e,
    \qquad
    B_p=\prod_{e\in\partial p}Z_e.
\end{equation*}
For definiteness, let us consider single-link $Z$ errors. A $Z$
operator on a link $e=\langle vv'\rangle$ anticommutes with the
two star stabilizers located at the endpoints $v$ and $v'$.
Following the construction in Sec.~\ref{sec21}, we assign an Ising spin
$\sigma_v=\pm1$ to each $X$-type stabilizer.
The classical interaction associated with the physical link $e$
is then
\begin{equation}
    b_e(\bm\sigma)
    =
    \sigma_v\sigma_{v'}.
    \label{eq:toric-bond}
\end{equation}
Thus, the classical model associated with this error sector is the
ordinary nearest-neighbor Ising model,
\begin{equation}
    H_{\rm Ising}[\bm\sigma]
    =
    -J\sum_{\langle vv'\rangle}
    \sigma_v\sigma_{v'}.
    \label{eq:clean-ising}
\end{equation}
Possible global relations among the star stabilizers only produce
an overall redundancy and will not affect the following local
argument.

We now regard the Boltzmann distribution of the clean model as a
prior distribution,
\begin{equation}
    P_{\beta}(\bm\sigma)
    =
    \frac{1}{Z_{\beta}}
    \exp\left[
        \beta J\sum_e b_e(\bm\sigma)
    \right].
    \label{eq:ising-prior}
\end{equation}
In the quantum problem, the parameter $\beta$ can arise from a
wavefunction deformation of the corresponding toric code state.
We perform a weak Gaussian measurement of every bond variable
$b_e=\pm1$. For a fixed spin configuration~$\bm\sigma$, the
conditional probability for obtaining a continuous measurement
outcome~$m_e$ is
\begin{equation}
    P_{\Delta}(m_e|\bm\sigma)
    \propto
    \exp\left[
        -\frac{\Delta}{2}
        \left(m_e-b_e(\bm\sigma)\right)^2
    \right].
    \label{eq:gaussian-measurement-ising}
\end{equation}
The measurement outcomes therefore define a classical Bayesian
inference problem. By Bayes' theorem, the posterior distribution
is
\begin{equation}
    P(\bm\sigma|\bm m)
    =
    \frac{1}{Z[\bm m]}
    \exp\left[
        \beta J\sum_e b_e(\bm\sigma)
        +
        \Delta\sum_e m_e b_e(\bm\sigma)
    \right],
    \label{eq:ising-posterior}
\end{equation}
where terms depending only on $\bm m$ have been absorbed into the
normalization.

To average products of posterior expectation values over
measurement records, we introduce $R$ replicas. After integrating over the
Gaussian measurement outcomes, the replicated model is
\begin{align}
    -H_R
    &=
    \beta J
    \sum_e\sum_{a=1}^{R}
    b_e(\bm\sigma^{(a)})
    +
    \frac{\Delta}{2}
    \sum_e
    \sum_{\substack{a,b=1\\a\neq b}}^{R}
    b_e(\bm\sigma^{(a)})
    b_e(\bm\sigma^{(b)}),
    \label{eq:toric-replica}
\end{align}
up to an additive constant. The physical learning problem is
obtained in the replica limit $R\to1$.
A special simplification occurs on the line, $\Delta=\beta J$.
On this line, \eqref{eq:toric-replica} can be written as
\begin{equation}
    -H_R
    =
    \frac{\Delta}{2}
    \sum_e
    \left[
        1+
        \sum_{a=1}^{R}
        b_e(\bm\sigma^{(a)})
    \right]^2
    +{\rm const}.
    \label{eq:toric-square-form}
\end{equation}

To expose the enhanced symmetry, introduce arbitrary Ising
variables $s_v=\pm1$ and perform the change of variables
\begin{equation}
    \widetilde{\sigma}^{(a)}_v
    =
    s_v\sigma^{(a)}_v.
\end{equation}
Since
\begin{equation*}
    b_e(\widetilde{\bm\sigma}^{(a)})
    =
    s_vs_{v'}
    b_e(\bm\sigma^{(a)}),
\end{equation*}
we have
\begin{equation}
    \left[
        1+\sum_{a=1}^{R}
        b_e(\widetilde{\bm\sigma}^{(a)})
    \right]^2
    =
    \left[
        s_vs_{v'}
        +
        \sum_{a=1}^{R}
        b_e(\bm\sigma^{(a)})
    \right]^2.
\end{equation}
We may therefore promote the auxiliary variables to an additional
replica,
i.e., 
   $ \sigma_v^{(R+1)}:=s_v.$
After summing over $\{s_v\}$, the partition function becomes
\begin{equation}
    Z_R
    =
    \frac{1}{2^{N_v}}
    \sum_{\{\bm\sigma^{(a)}\}_{a=1}^{R+1}}
    \exp\left[
        \frac{\Delta}{2}
        \sum_e
        \left(
            \sum_{a=1}^{R+1}
            b_e(\bm\sigma^{(a)})
        \right)^2
    \right].
    \label{eq:toric-Rplus1}
\end{equation}
The formulation in~\eqref{eq:toric-Rplus1} has an enlarged
replica permutation symmetry $S_{R+1}$. In addition, it has a
local gauge invariance
\begin{equation}
    \sigma_v^{(a)}
    \longrightarrow
    \eta_v\sigma_v^{(a)},
    \qquad
    a=1,\ldots,R+1,
    \qquad
    \eta_v=\pm1.
    \label{eq:toric-replica-gauge}
\end{equation}
Indeed, every $b_e(\bm\sigma^{(a)})$ acquires the same factor
$\eta_v\eta_{v'}$ for all replicas, which disappears upon
squaring the replica sum. The line~$\Delta=\beta J$ is therefore a higher
Nishimori line: the $R\to1$ learning problem admits an equivalent
gauge invariant formulation with $R+1\to2$ replicas and enlarged
$S_{R+1}$ symmetry~\cite{patil2026higher}.

As a first consequence, consider the spin correlation
\begin{equation}
    O_{vv'}(\bm\sigma)
    =
    \sigma_v\sigma_{v'}.
\end{equation}
On the higher Nishimori line, the standard Nishimori argument
gives
\begin{equation}
    \overline{
        \left\langle O_{vv'}\right\rangle_{\bm m}^{\,2q-1}
    }
    =
    \overline{
        \left\langle O_{vv'}\right\rangle_{\bm m}^{\,2q}
    },
    \qquad
    q=1,2,\ldots.
    \label{eq:toric-odd-even}
\end{equation}
In particular,
\begin{equation}
    \overline{
        \left\langle O_{vv'}\right\rangle_{\bm m}^{\,2}
    }
    =
    \overline{
        \left\langle O_{vv'}\right\rangle_{\bm m}
    }.
    \label{eq:toric-EA-first}
\end{equation}
Furthermore, the law of total expectation gives
\begin{align}
    \overline{
        \left\langle O_{vv'}\right\rangle_{\bm m}
    }
=
    \int D\bm m\,P(\bm m)
    \sum_{\bm\sigma}
    P(\bm\sigma|\bm m)O_{vv'}(\bm\sigma)
   =
    \sum_{\bm\sigma}
    P_{\beta}(\bm\sigma)O_{vv'}(\bm\sigma)
    =
    \left\langle
        \sigma_v\sigma_{v'}
    \right\rangle_{\rm clean}.
\end{align}
Consequently,
\begin{equation}
    \overline{
        \left\langle
            \sigma_v\sigma_{v'}
        \right\rangle_{\bm m}^{\,2}
    }
    =
    \left\langle
        \sigma_v\sigma_{v'}
    \right\rangle_{\rm clean}.
    \label{eq:toric-EA-clean}
\end{equation}
For the ordinary square-lattice Ising model, this reproduces the
higher Nishimori identity discussed in~\cite{patil2026higher}.

\subsection{Rank-2 toric code and dipole Ising model}\label{222}
\begin{figure}
    \centering
    \includegraphics[width=0.95\linewidth]{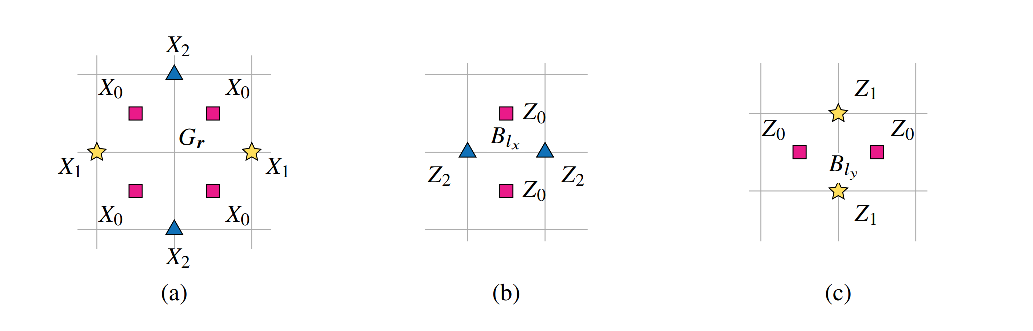}
    \caption{Stabilizers of the rank-2 toric code~\eqref{hami1}.  The yellow star, blue triangle, and red square represent Pauli operators
acting on qubits of types $1$, $2$, and $0$, respectively.}
    \label{fig:placeholder}
\end{figure}

We next consider the $\mathbb Z_2$ rank-2 toric code~\cite{PhysRevB.106.045145}~\footnote{See also, e.g.,~\cite{PhysRevB.110.045146,kim2026transparent} for quantum information aspects of a model with the modulated symmetry.}. We first introduce the model and its loop operators, and then apply the identities derived above.
\subsubsection{Model and loop operators}
The model contains three qubits, labeled by $a=0,1,2$ per unit
cell.  The type-$0$ qubit is located at the center of each plaquette,
whereas the type-$1$ and type-$2$ qubits are located at the vertices.
There are three types of mutually commuting
stabilizers. The $X$-type
stabilizer is given by
\begin{align}
G_{\mathbf r}
=
X_{1,\mathbf r-\hat{\mathbf x}}
X_{1,\mathbf r+\hat{\mathbf x}}
X_{2,\mathbf r-\hat{\mathbf y}}
X_{2,\mathbf r+\hat{\mathbf y}}
\prod_{\epsilon_x,\epsilon_y=\pm1}
X_{0,\mathbf r+
(\epsilon_x\hat{\mathbf x}+\epsilon_y\hat{\mathbf y})/2}.
\label{eq:R2TC-a}
\end{align}
The two $Z$-type stabilizers can be written as
\begin{align}
B_{\mathbf{l}_x}
&=
Z_{2,\bm r}
Z_{2,\bm r+\hhx}
Z_{0,\bm r+\hhx/2+\hhy/2}
Z_{0,\bm r+\hhx/2-\hat y/2},\nonumber
\\
B_{\mathbf{l}_y}
&=
Z_{1,\bm r}
Z_{1,\bm r+\hhy}
Z_{0,\bm r+\hhx/2+\hhy/2}
Z_{0,\bm r-\hat x/2+\hhy/2},\label{bx}
\end{align}
where $\mathbf{l}_x=\br+\hhx/2$,  $\mathbf{l}_y=\br+\hhy/2$. 
The Hamiltonian is defined by
\begin{eqnarray}
    H=-\sum_\br G_\br-\sum_{\mathbf{l}_x}B_{\mathbf{l}_x}-\sum_{\mathbf{l}_y}B_{\mathbf{l}_y}\label{hami1}.
\end{eqnarray}
See also Fig.~\ref{fig:placeholder}.
%
%
\par
\begin{figure}
    \centering
    \includegraphics[width=0.9\linewidth]{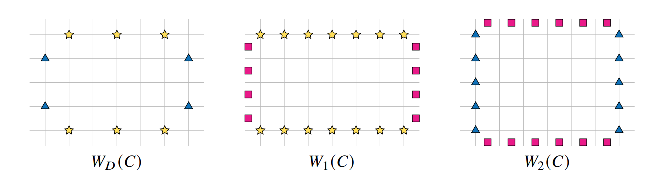}
    \caption{Three types of loop operators defined in~\eqref{eq:r2tc-loop-operators}. The three loops are related via lattice translation as indicated in~\eqref{eq:R2TC-translation-loops}. The yellow star, blue triangle, and red square represent Pauli $Z_1$, $Z_2$, and $Z_0$, respectively. }
    \label{fig:}
\end{figure}

This model admits three types of loop operators consisting of Pauli $Z$ operators (Fig.~\ref{fig:}).
Let \(C\) be the rectangular contour with corners
$(x_L,y_B)$, $(x_R,y_B)$, $(x_R,y_T)$, and
$(x_L,y_T)$, where
\begin{equation*}
      x_L=x_0,\qquad x_R=x_0+2r_x,\qquad
  y_B=y_0,\qquad y_T=y_0+2r_y .
\end{equation*}
and \(x_0,y_0\in 2\mathbb Z\). The three loop operators supported on
and immediately adjacent to \(C\) are defined as
\begin{eqnarray}
W_D(C)&=&
 \prod_{\hx=x_L}^{x_R}\bigl(Z_{1,(\hx,y_B)}\bigr)^{\hx}
 \times\prod_{\hy=y_B}^{y_T}\bigl(Z_{2,(x_R,\hy)}\bigr)^{\hy}
 \times\prod_{\hx=x_L}^{x_R}\bigl(Z_{1,(\hx,y_T)}\bigr)^{\hx}\times
 \prod_{\hy=y_B}^{y_T}\bigl(Z_{2,(x_L,\hy)}\bigr)^{\hy},
 \nonumber\\
W_1(C)&=&
 \prod_{\hx=x_L}^{x_R}\left[ Z_{1,(\hx,y_B)}Z_{1,(\hx,y_T)}\right]
 \times\prod_{\hy=y_B}^{y_T-1}\left[
 Z_{0,(x_L-\frac12,\hy+\frac12)}
 Z_{0,(x_R+\frac12,\hy+\frac12)}\right],
 \nonumber\\
W_2(C)&=&
 \prod_{\hy=y_B}^{y_T} \left[Z_{2,(x_L,\hy)}Z_{2,(x_R,\hy)}\right]
 \times\prod_{\hx=x_L}^{x_R-1}\left[
 Z_{0,(\hx+\frac12,y_B-\frac12)}
 Z_{0,(\hx+\frac12,y_T+\frac12)}\right].
 \label{eq:r2tc-loop-operators}
\end{eqnarray}
Here, the exponents in \(W_D\) are understood modulo two.

These loops are related via lattice translation. Indeed, within the fluxless subspace, $B_{\mathbf{l}_x}=B_{\mathbf{l}_y}=+1$, after some algebra, one finds~\cite{PhysRevB.111.035149}
\begin{align}
T_x W_D(C)T_x^{-1}
&=
W_D(C)W_1(C),
\nonumber\\
T_y W_D(C)T_y^{-1}
&=
W_D(C)W_2(C).
\label{eq:R2TC-translation-loops}
\end{align}
Equivalently, using $W_D^2=1$,
\begin{align*}
W_1(C)
&=
W_D(C)\,T_xW_D(C)T_x^{-1},
\nonumber\\
W_2(C)
&=
W_D(C)\,T_yW_D(C)T_y^{-1}.
\end{align*}
Note that equality holds after projection onto the common \(+1\) eigenspace of all \(Z\)-type stabilizers.
Hence, the product of two translated modulated dipole loops gives
a homogeneous charge loop.
Relations in~\eqref{eq:R2TC-translation-loops} are the loop-operator
analogue of the transformation of a dipole moment under
translations.  For
\begin{equation*}
D_i=\int \mathrm{d}^2\bm r\,r_i\rho(\bm r),
\qquad
Q=\int \mathrm{d}^2\bm r\,\rho(\bm r),
\end{equation*}
a translation by one lattice spacing changes $D_i$ by the total
charge $Q$.  Correspondingly, a lattice translation transforms
the dipole loop into itself multiplied by a homogeneous charge
loop.  We therefore refer to $W_D$ as the dipole loop and to
$W_1$ and $W_2$ as the associated homogeneous charge, or
monopole, loops~\cite{Ebisu:2023idd,Pace:2025hpb,10.21468/SciPostPhys.20.4.117}.
\subsubsection{Learning problem}\label{444}
We now formulate the learning problem for this code.
Based on the argument in Sec.~\ref{sec21}, 
a single $Z_{1,\bm r}$ error anticommutes with
$G_{\bm r-\hat x}$ and $G_{\bm r+\hhx}$, and therefore gives
the classical interaction
\begin{equation}
B_{1,\bm r}
=
\sigma_{\bm r-\hat x}\sigma_{\bm r+\hhx}.\label{103}
\end{equation}
Likewise, a single $Z_{2,\bm r}$ error anticommutes with
$G_{\bm r-\hat y}$ and $G_{\bm r+\hhy}$, giving
the classical interaction
\begin{equation}
B_{2,\bm r}
=
\sigma_{\bm r-\hat y}\sigma_{\bm r+\hhy}.\label{104}
\end{equation}
Finally, for a type-$0$ qubit on the plaquette 
$Z_{0,\mathbf r+(\hat x+\hat y)/2}$ anticommutes with
$G_{\mathbf r}$, $G_{\mathbf r+\hat x}$,
$G_{\mathbf r+\hat y}$, and $G_{\mathbf r+\hat x+\hat y}$ which gives the following classical interaction:
\begin{equation}
B_{0,\bm r}
=
\sigma_{\bm r}
\sigma_{\bm r+\hhx}
\sigma_{\bm r+\hhy}
\sigma_{\bm r+\hhx+\hhy}.
\end{equation}
Thus, up to a shift of the lattice coordinates, the statistical
model associated with the $Z$-error sector of the rank-2 toric code is
\begin{align}
H_{\rm dip}
=
-\sum_{\bm r}
\Big[
&
J_x\sigma_{\bm r}\sigma_{\bm r+2\hhx}
+
J_y\sigma_{\bm r}\sigma_{\bm r+2\hhy}
+
J_\square
\sigma_{\bm r}
\sigma_{\bm r+\hhx}
\sigma_{\bm r+\hhy}
\sigma_{\bm r+\hhx+\hhy}
\Big].
\label{eq:R2TC-dipole-Ising}
\end{align}
The stabilizer relations of the rank-2 toric code become spatially modulated symmetries of the associated classical model. Explicitly, assuming the infinite lattice system, 
the Hamiltonian~\eqref{eq:R2TC-dipole-Ising} respects the following modulated symmetry:
\begin{eqnarray}
    \sigma_\br\to\eta\zeta^{\hx}_x\zeta^{\hy}_y\sigma_\br,\quad \eta,\zeta_x,\zeta_y=\pm1.
\end{eqnarray}
Setting $\zeta_x=-1$ or $\zeta_y=-1$, with the other parameters
equal to $+1$, generates the $x$- or $y$-modulated
$\mathbb Z_2$ transformation, respectively.
Setting $\eta=-1$, with $\zeta_x=\zeta_y=+1$, gives the uniform
spin flip.

We weakly measure the three types of qubits of the rank-2 toric code with
Gaussian measurement strengths~$\Delta_x,\Delta_y,\Delta_{\square}$. After averaging over the
measurement outcomes, the $R$-replica theory takes the form
\begin{align}
    -H_R
    =
    \sum_{\bm r}
    \sum_{\mu=x,y,\square}
    \Bigg[
        K_{\mu}
        \sum_{a=1}^{R}
        b_{\bm r}^{\mu,(a)}
        +
        \frac{\Delta_{\mu}}{2}
        \sum_{\substack{a,b=1\\a\neq b}}^{R}
        b_{\bm r}^{\mu,(a)}
        b_{\bm r}^{\mu,(b)}
    \Bigg],
    \label{eq:dipole-replica}
\end{align}
where $K_{\mu}:=\beta J_{\mu}.$

The higher Nishimori condition is now a manifold in the coupling
space,
\begin{equation}
     \Delta_{\mu}=K_{\mu},
    \qquad
    \mu=x,y,\square.
  \label{eq:dipole-HN}
\end{equation}
On this manifold, we have
\begin{equation}
    -H_R
    =
    \frac{1}{2}
    \sum_{\bm r,\mu}
    K_{\mu}
    \left[
        1+
        \sum_{a=1}^{R}
        b_{\bm r}^{\mu,(a)}
    \right]^2
    +{\rm const}.
    \label{eq:dipole-square}
\end{equation}

The key observation is that all three interactions satisfy the
same multiplicative property. For an arbitrary Ising
configuration $\bm s$,
\begin{align}
    b_{\bm r}^{x}(\bm\sigma\bm s)
    =
    b_{\bm r}^{x}(\bm\sigma)
    b_{\bm r}^{x}(\bm s),
    \label{eq:dipole-multiplicative}
\end{align}
where
$(\bm\sigma\bm s)_{\bm r}:=\sigma_{\bm r}s_{\bm r}$. The same property holds for $b_{\bm r}^{y}(\bm\sigma\bm s)$ and $b_{\bm r}^{\square}(\bm\sigma\bm s)$.
Consequently, precisely the same auxiliary-spin construction as
in the ordinary Ising model promotes $\bm s$ to an $(R+1)$th
replica. We obtain
\begin{equation}
    Z_R
    =
    \frac{1}{2^{N_s}}
    \sum_{\{\bm\sigma^{(a)}\}_{a=1}^{R+1}}
    \exp\left[
        \frac{1}{2}
        \sum_{\bm r,\mu}
        K_{\mu}
        \left(
            \sum_{a=1}^{R+1}
            b_{\bm r}^{\mu}(\bm\sigma^{(a)})
        \right)^2
    \right].
    \label{eq:dipole-Rplus1}
\end{equation}
This formulation again has $S_{R+1}$ replica permutation
symmetry as well as position dependent local gauge invariance
\begin{equation}
    \sigma_{\bm r}^{(a)}
    \longrightarrow
    \eta_{\bm r}\sigma_{\bm r}^{(a)},
    \qquad
    a=1,\ldots,R+1.
    \label{eq:dipole-gauge}
\end{equation}
with $\eta_{\br}=\pm1$. On the higher Nishimori manifold, the replicated theory possesses an enlarged position-dependent gauge redundancy. 

We now introduce the three learning correlators associated with
the three loop operators of the rank-2 toric code. To do so, we think of an open string of the dipole loop $W_D$ which has the following form
\begin{eqnarray}
     \prod_{\hx=x_L}^{x_R}\bigl(Z_{1,(\hx,y_B)}\bigr)^{\hx}
 \times\prod_{\hy=y_B}^{y_T}\bigl(Z_{2,(x_R,\hy)}\bigr)^{\hy}.\label{qew}
\end{eqnarray}
A single $Z_{1,\br}$ operator maps to the interaction  $B_{1,\br}$
in~\eqref{103}, while  $Z_{2,\br}$ maps to~$B_{2,\br}$ in~\eqref{104}. 
Consequently, the classical image of the open dipole loop~\eqref{qew} is described by
\begin{eqnarray}
     \prod_{\hx=x_L}^{x_R}\bigl(B_{1,(\hx,y_B)}\bigr)^{\hx}
 \times\prod_{\hy=y_B}^{y_T}\bigl(B_{2,(x_R,\hy)}\bigr)^{\hy}=\sigma_{(x_0,y_B)}\sigma_{(x_R,y_T)}.
\end{eqnarray}
More generally, we write such an interaction as
\begin{equation}
O_D(\bm r,\bm r')
:=
\sigma_{\bm r}\sigma_{\bm r'}.
\label{eq:R2TC-OD}
\end{equation}
with
$\bm r-\bm r'\in(2\mathbb Z)^2\; (|\br-\br^\prime|\gg 1),$
so that $\bm r$ and $\bm r'$ belong to the same parity
sublattice. Likewise, we define classical spin interactions corresponding to open loops $W_1$ and $W_2$ as
\begin{align}
O_1(\bm r,\bm r')
&:=
\sigma_{\bm r}
\sigma_{\bm r+\hhx}
\sigma_{\bm r'}
\sigma_{\bm r'+\hhx},
\label{eq:R2TC-O1}
\\
O_2(\bm r,\bm r')
&:=
\sigma_{\bm r}
\sigma_{\bm r+\hhy}
\sigma_{\bm r'}
\sigma_{\bm r'+\hhy}.
\label{eq:R2TC-O2}
\end{align}
Note that the following relation holds
\begin{eqnarray}
    O_{a}(\br,\br^\prime)= O_{D}(\br,\br^\prime)O_D^{(a)}(\br,\br^\prime)~(a=1,2)
\end{eqnarray}
where
\begin{equation}
O_D^{(a)}(\bm r,\bm r')
:=
\sigma_{\bm r+\bm e_a}
\sigma_{\bm r'+\bm e_a},
\qquad a=1,2.
\end{equation}
which is obtained by
translation of $O_D(\br,\br^\prime)$ in the $x$- or $y$-direction. 
Using these interactions, we further define their EA-type learning correlators by
\begin{align}
\mathcal C_D(\bm r,\bm r')
&:=
\overline{
\left\langle
O_D(\bm r,\bm r')
\right\rangle_{\bm m}^{2}
},\quad 
\mathcal C_1(\bm r,\bm r')
:=
\overline{
\left\langle
O_1(\bm r,\bm r')
\right\rangle_{\bm m}^{2}
},\quad 
\mathcal C_2(\bm r,\bm r'):=
\overline{
\left\langle
O_2(\bm r,\bm r')
\right\rangle_{\bm m}^{2}
}.
\label{eq:R2TC-learning-correlators}
\end{align}
On the higher Nishimori manifold, the generalized Nishimori
identity gives
\begin{equation}
\mathcal C_\alpha(\bm r,\bm r')
=
\left\langle
O_\alpha(\bm r,\bm r')
\right\rangle_{\rm clean},
\qquad
\alpha=D,1,2.
\label{eq:R2TC-clean-correlators}
\end{equation}
Because the clean dipole Ising model has nonnegative
ferromagnetic couplings, the GKS inequality
can be applied to
$O_D$ and its translated copies. For $a=1,2$, it gives
\begin{align}
\mathcal C_a(\bm r,\bm r')
&=
\left\langle
O_D(\bm r,\bm r')
O_D^{(a)}(\bm r,\bm r')
\right\rangle_{\rm clean}
\geq
\left\langle
O_D(\bm r,\bm r')
\right\rangle_{\rm clean}
\left\langle
O_D^{(a)}(\bm r,\bm r')
\right\rangle_{\rm clean}.
\end{align}
Translation invariance implies
\begin{equation}
\left\langle
O_D^{(a)}(\bm r,\bm r')
\right\rangle_{\rm clean}
=
\left\langle
O_D(\bm r,\bm r')
\right\rangle_{\rm clean}.
\end{equation}
Consequently,
\begin{equation}
\boxed{
\mathcal C_a(\bm r,\bm r')
\geq
\left[
\mathcal C_D(\bm r,\bm r')
\right]^2,
\qquad a=1,2.
}
\label{eq:R2TC-exact-hierarchy}
\end{equation}
This is the exact
dipole--monopole hierarchy for the rank-2 toric code learning
problem. In particular, long range learnability of the
spatially modulated dipole loop $W_D$ necessarily implies
long range learnability of both homogeneous monopole loops
$W_1$ and $W_2$. The converse implication does not follow from
the inequality: the homogeneous loops may remain learnable even
when the dipole loop is not.
If the correlators exhibit power law scaling,
\begin{equation}
     \mathcal C_D(r)\sim A_D r^{-2h_D},
 \qquad
 \mathcal C_a(r)\sim A_a r^{-2h_a},
 \qquad a=1,2,
\end{equation}
with positive amplitudes, the bound~\eqref{eq:R2TC-exact-hierarchy} implies $h_a\leq 2h_D$.
This explicit relation hints at the possibility of diagnosing and constraining the criticality of general learning problems in future studies. In previous works on learning problems~\cite{7dpt-d4s5,patil2026higher,Patil:2026potts}, conformal field theory (CFT) descriptions have been established on the higher Nishimori manifold. 
It would be interesting to investigate whether we have criticality of the present learning problem. We come back to this issue for a future work. 

In Appendix~\ref{xpaul}, we discuss the corresponding learning problem with
weak measurements of Pauli~$X$ operators.
In this case, we have the complementary hierarchy: 
two distinct modulated parent loops share a common homogeneous loop. This reversal relative to the Pauli-$Z$ measurement case reflects the opposite action of lattice translations on the electric and magnetic loop operators of the rank-2 toric code~\cite{PhysRevB.111.035149}.
Also, in Appendix~\ref{xcube}, we apply our
results to the X-cube model~\cite{Vijay}, whose classical representation possesses
planar subsystem symmetries.

\section{Learning problem in finite group quantum doubles}
\label{sec:qd-electric-learning}\label{QD}
So far, we have studied the learning problem in \emph{Abelian}
topological stabilizer codes. In this section, we extend the discussion
to  quantum double $D(G)$ for a finite group
$G$~\cite{KITAEV20032}, which may admit \emph{non-Abelian} anyons~\footnote{See also, e.g.,~\cite{PhysRevA.69.032306,PhysRevB.78.115421,Wootton:2010dah,Cui:2019lvb,bravyi2022adaptive} for earlier expositions of this model in view of quantum information 
and~\cite{5ywn-6d3q,jing2026integer,Vadali:2026,sala2026climbing} for recent studies of the model under decoherence.}. 
The pure electric charges are labeled by irreducible representations
of $G$ and can be non-Abelian when $G$ is non-Abelian.
We consider Gaussian weak measurements of local operators that create
pairs of these charges. We first establish an exact identity between the EA correlators of open
electric strings and the corresponding clean correlators.
Under an additional positivity condition on the classical interactions,
this identity leads to inequalities relating the learnability of
different electric charges.
\subsection{Quantum double model and weak measurements}
\label{subsec:qd-model-measurement}

We consider an oriented square lattice with a finite, connected set of
vertices $V$ and edges~$E$. Each oriented edge $e$ carries a
$|G|$-dimensional Hilbert space. The many-body Hilbert space is given by
\begin{equation}
 \mathcal H_e=\operatorname{span}\{|u\rangle_e:u\in G\},
 \qquad \mathcal H=\bigotimes_e\mathcal H_e.
 \label{eq:qd-Hilbert}
\end{equation}
For $h\in G$, we define $L_h|u\rangle_e=|hu\rangle_e$ and
$R_h|u\rangle_e=|uh\rangle_e$. The operator $A_v(h)$ acts with $L_h$ on edges
oriented away from $v$ and with $R_{h^{-1}}$ on edges oriented toward
$v$. The vertex projector is $A_v=|G|^{-1}\sum_{h\in G}A_v(h)$.
The plaquette projector $B_p$ imposes trivial holonomy around each
plaquette. Since $G$ need not be Abelian, we multiply the edge
variables in the order along $\partial p$:
\begin{equation}
 B_p|\{u_e\}\rangle
 =\delta_G\!\left(\overrightarrow{\prod}_{e\in\partial p}
  u_e^{\epsilon(p,e)},1_G\right)|\{u_e\}\rangle .
\end{equation}
Here, $1_G$ denotes the identity element of $G$, and
$\epsilon(p,e)=\pm1$ specifies whether the orientation of $e$ agrees with that of
the plaquette boundary. Also, $\delta_G(a,b)$ is one for $a=b$ and zero otherwise.
The Hamiltonian is given by
\begin{equation}
 H_{D(G)}=-\sum_v A_v-\sum_p B_p.
\end{equation}
An anyon of the quantum double $D(G)$ is labeled by a pair
$([g],\pi)$, where $[g]$ is a conjugacy class of $G$ and
$\pi\in\operatorname{Irr}(C_G(g))$ is an irreducible representation of the centralizer
$C_G(g)=\{h\in G:hg=gh\}$~\cite{KITAEV20032}. The pure electric sector has trivial flux, $[g]=[1_G]$.
Since $C_G(1_G)=G$, a pure electric charge is labeled by $( [1_G],\rho)$ with $\rho\in\operatorname{Irr}(G)$.
In the following, we omit the trivial flux label and denote pure
electric charges by $\rho$. Their fusion rules are given by
\begin{equation}
  ([1_G],\lambda)\otimes([1_G],\eta)
    =\bigoplus_{\nu\in\operatorname{Irr}(G)}
      N_{\lambda\eta}^{\nu}([1_G],\nu),\qquad
 N_{\lambda\eta}^{\nu}=\dim\operatorname{Hom}_G(\nu,\lambda\otimes\eta).
\end{equation}
\par
We work in the zero-flux sector with trivial global holonomies.
Introducing a group element $g_i\in G$ at each vertex, we write the
physical edge configurations as
\begin{equation}
 u_{ij}=g_i g_j^{-1}
 \qquad (e=(i,j)\text{ oriented from }i\text{ to }j).
 \label{eq:qd-puregauge}
\end{equation}
The simultaneous right multiplication $g_i\mapsto g_i k$ leaves all edge variables
unchanged. Let $\mathcal C$ denote the set of distinct pure-gauge edge
configurations, and write $|\boldsymbol u\rangle=\bigotimes_e|u_e\rangle_e$ for the corresponding many-body basis
state. The normalized ground state in this sector is the equal-amplitude
superposition
\begin{equation}
 |\Psi_0\rangle=\frac{1}{\sqrt{|\mathcal C|}}
 \sum_{\boldsymbol u\in\mathcal C}|\boldsymbol u\rangle.
 \label{eq:qd-groundstate}
\end{equation}
For later convenience, we define the normalized sum over the vertex
variables as
\begin{equation}
 \int\!\mathcal Dg\,F(\boldsymbol g)
 :=\frac{1}{|G|^{|V|}}\sum_{\boldsymbol g\in G^{V}}F(\boldsymbol g).
 \label{eq:qd-uniform-average}
\end{equation}
On a connected graph, exactly $|G|$ choices of the vertex variables
represent each pure-gauge edge configuration. These choices are related
by $g_i\mapsto g_i k$, and the resulting factor cancels in normalized averages of
functions of the edge variables. In particular, $|\mathcal C|=|G|^{|V|-1}$.

Let $\rho\in\operatorname{Irr}(G)$ be a unitary irreducible representation of dimension $d_\rho$,
with representation matrix~$D^\rho(u)$. We define the following diagonal
operators on a single edge:
\begin{equation}
 \widehat T^{\rho}_{ab,e}|u\rangle_e
 =\frac{D^\rho_{ab}(u)}{\sqrt{d_\rho}}|u\rangle_e,
 \qquad a,b=1,\ldots,d_\rho.
 \label{eq:qd-electric-operator}
\end{equation}
When acting on the ground state, these operators create
electric charge--anticharge pairs at the endpoints of the edge.
The indices $a,b$ label the representation components at the two
endpoints. Note that these operators are not projectors onto a definite
charge sector.

We weakly measure their real and imaginary parts. To this end, we
collect all matrix components into the real vector~\footnote{When $D^\rho(u)$ is real for every $u\in G$, the imaginary components
vanish. They may be retained as Gaussian variables carrying no information,
or omitted with a corresponding reduction of $n_\rho$.}
\begin{equation}
 \boldsymbol f_\rho(u):=\frac{1}{\sqrt{d_\rho}}
 \bigl(\operatorname{Re}D^\rho_{ab}(u),
       \operatorname{Im}D^\rho_{ab}(u)\bigr)_{a,b=1}^{d_\rho}
 \in\mathbb R^{n_\rho},\qquad n_\rho=2d_\rho^2.
 \label{eq:qd-f-vector}
\end{equation}
The index $A=1,\ldots,n_\rho$ labels a matrix element $(a,b)$
together with its real or imaginary component.
We use the Euclidean inner product $x\cdot y=\sum_A x_Ay_A$ and the
squared norm $\|x\|^2=\sum_A x_A^2$.
From the unitarity of $D^\rho$, we obtain
%
\begin{align}
 \boldsymbol f_\rho(u)\cdot\boldsymbol f_\rho(v)
 &=\frac{1}{d_\rho}\operatorname{Re}\sum_{a,b}
       D^\rho_{ab}(u)D^\rho_{ab}(v)^*=\frac{1}{d_\rho}\operatorname{Re}
       \operatorname{Tr}\bigl[D^\rho(u)D^\rho(v)^\dagger\bigr]
 =\frac{\operatorname{Re}\chi_\rho(uv^{-1})}{d_\rho}.
 \label{eq:qd-feature-identity}
\end{align}
In particular, we have
\begin{equation}
 \|\boldsymbol f_\rho(u)\|^2=1,
 \qquad c_\rho(u):=\boldsymbol f_\rho(1_G)\cdot\boldsymbol f_\rho(u)
 =\frac{\operatorname{Re}\chi_\rho(u)}{d_\rho}.
 \label{eq:qd-c-and-norm}
\end{equation}

We choose a set $\mathcal S\subseteq\operatorname{Irr}(G)$ of nontrivial representations to be measured.
For each pair $(e,\rho)$, the measurement record is a real vector $\boldsymbol m_{e,\rho}\in\mathbb R^{n_\rho}$.
All components are measured with the same Gaussian precision $\Delta_{e,\rho}>0$,
which may depend on the edge and the representation.
The corresponding Hermitian operators are given by
\begin{equation}
 \widehat f_{\rho,A}(\widehat u_e)
 :=\sum_{u\in G} f_{\rho,A}(u)|u\rangle_e\langle u|.
\end{equation}
These operators are diagonal in the group basis and therefore commute.
The joint Kraus operator is the product of the Gaussian Kraus operators
for the individual components:
\begin{align}
 M_{e,\rho}(\boldsymbol m)
 =\prod_{A=1}^{n_\rho}
 \left(\frac{\Delta_{e,\rho}}{2\pi}\right)^{1/4}
 \exp\!\left[-\frac{\Delta_{e,\rho}}4
       (m_A-\widehat f_{\rho,A})^2\right]=\left(\frac{\Delta_{e,\rho}}{2\pi}\right)^{n_\rho/4}
 \exp\!\left[-\frac{\Delta_{e,\rho}}4
       \sum_{A=1}^{n_\rho}(m_A-\widehat f_{\rho,A})^2\right].
 \label{eq:qd-kraus}
\end{align}
The sum in the exponent is the squared Euclidean norm $\|\boldsymbol m-\widehat{\boldsymbol f}_\rho\|^2$, taken
componentwise. For a fixed group basis state $|u\rangle_e$, we have $\widehat f_{\rho,A}|u\rangle_e=f_{\rho,A}(u)|u\rangle_e$.
The Born rule then gives the conditional probability density for the
measurement record:
\begin{align}
 p_\Delta(\boldsymbol m\mid u)
 &=\langle u|M_{e,\rho}(\boldsymbol m)^\dagger
                         M_{e,\rho}(\boldsymbol m)|u\rangle_e=\left(\frac{\Delta_{e,\rho}}{2\pi}\right)^{n_\rho/2}
 \exp\!\left[-\frac{\Delta_{e,\rho}}2
       \|\boldsymbol m-\boldsymbol f_\rho(u)\|^2\right].
 \label{eq:qd-likelihood}
\end{align}
For fixed $u$, the record components are independent Gaussian
variables with means $f_{\rho,A}(u)$ and variances $1/\Delta_{e,\rho}$.
Thus, $\int d^{n_\rho}m\,p_\Delta(\boldsymbol m\mid u)=1$ for every $u$, and the Kraus operators satisfy the
completeness relation $\int d^{n_\rho}m\,M_{e,\rho}^\dagger M_{e,\rho}=I$.

As in the CSS code examples, we deform the initial state before
applying these measurements:
\begin{equation}
 |\Psi_K\rangle=\frac{1}{\sqrt{Z_K}}
 \exp\!\left[\frac12\sum_{e,\rho\in\mathcal S}
 K_{e,\rho}c_\rho(\widehat u_e)\right]|\Psi_0\rangle,
 \quad Z_K=\int\!\mathcal Dg\,
 \exp\!\left[\sum_{e,\rho\in\mathcal S}
 K_{e,\rho}c_\rho(u_e)\right].
 \label{eq:qd-deformed-state}
\end{equation}
The diagonal probabilities of this state define the \emph{clean prior}
\begin{equation}
 P_K(\boldsymbol g)=\frac{1}{Z_K}
 \exp\!\left[\sum_{e,\rho\in\mathcal S}
 K_{e,\rho}c_\rho(u_e)\right],
 \qquad\int\!\mathcal Dg\,P_K(\boldsymbol g)=1.
 \label{eq:qd-prior}
\end{equation}
Here, \emph{clean} means that the average is taken before conditioning
on a measurement record.

Let $\boldsymbol m=\{\boldsymbol m_{e,\rho}\}_{e,\rho\in\mathcal S}$ denote the complete measurement record. For a fixed edge
configuration, the likelihood is a product of the Gaussian factors
for the measured components. Using Bayes' rule and $\|\boldsymbol f_\rho(u)\|^2=1$, we obtain
\begin{align}
 P_{K,\Delta}(\boldsymbol g\mid\boldsymbol m)
 \propto P_K(\boldsymbol g)
 \prod_{e,\rho\in\mathcal S}
 p_\Delta(\boldsymbol m_{e,\rho}\mid u_e)
 \propto\exp\!\left[\sum_{e,\rho\in\mathcal S}
 \left\{K_{e,\rho}c_\rho(u_e)
 +\Delta_{e,\rho}\boldsymbol m_{e,\rho}\cdot
                         \boldsymbol f_\rho(u_e)\right\}\right].
 \label{eq:qd-posterior}
\end{align}
The posterior is normalized with respect to $\mathcal Dg$.
The Born probability density of the complete record is obtained by
averaging the likelihood over the prior:
\begin{equation*}
 p_{K,\Delta}(\boldsymbol m)
 =\int\!\mathcal Dg\,P_K(\boldsymbol g)
   \prod_{e,\rho\in\mathcal S}
   p_\Delta(\boldsymbol m_{e,\rho}\mid u_e).
\end{equation*}
Thus, the record distribution is determined by the deformed quantum
state, rather than by a Gaussian distribution chosen independently of
the state.

\subsection{Matrix EA correlators}
\label{subsec:qd-HN}

We next define the matrix EA correlator associated with an open
electric string. For an oriented path $\gamma:x\to y$, we multiply the edge
variables in path order, using $u_e^{-1}$ whenever the path runs opposite
to the orientation of $e$. In the pure-gauge sector, the product
reduces to $U_{xy}=g_xg_y^{-1}$. For an irreducible representation $\lambda$, we define
the open string matrix conditioned on the measurement record as
\begin{equation}
 A_\lambda(\boldsymbol m;x,y)
 :=\int\!\mathcal Dg\,
 P_{K,\Delta}(\boldsymbol g\mid\boldsymbol m)
 D^\lambda(g_xg_y^{-1}).
 \label{eq:qd-A}
\end{equation}
The corresponding matrix EA correlator is defined by
\begin{equation}
 Q_\lambda(K,\Delta;x,y)
 :=\int d\boldsymbol m\,p_{K,\Delta}(\boldsymbol m)
 \frac{\operatorname{Tr}\bigl[A_\lambda(\boldsymbol m)
          A_\lambda(\boldsymbol m)^\dagger\bigr]}{d_\lambda}.
 \label{eq:qd-EA}
\end{equation}
The matrix $A_\lambda$ is averaged over group configurations at a fixed
record $\boldsymbol m$. We then average its normalized squared Frobenius norm
over records with their Born probabilities. This generalizes the
squared spin--spin correlator of the $\mathbb Z_2$ toric code.

For comparison, we define the \emph{clean correlator} by averaging over
the same deformed prior without conditioning on a measurement record:
\begin{align}
 C_\lambda(K;x,y)
 :=\int\!\mathcal Dg\,P_K(\boldsymbol g)
        \frac{\chi_\lambda(g_xg_y^{-1})}{d_\lambda}=\frac{1}{Z_K}\int\!\mathcal Dg\,
 \exp\!\left[\sum_{e,\rho\in\mathcal S}
   K_{e,\rho}c_\rho(u_e)\right]
 \frac{\chi_\lambda(g_xg_y^{-1})}{d_\lambda}.
 \label{eq:qd-clean-correlator}
\end{align}
As a consistency check, for $G=\mathbb Z_2$, the nontrivial character gives
$C_\lambda(K;x,y)=\langle\sigma_x\sigma_y\rangle_K$. This is the ordinary Ising spin--spin correlator appearing in
the $\mathbb{Z}_2$ toric code result~\eqref{eq:toric-EA-clean}.
In general, $Q_\lambda(K,\Delta;x,y)$ and $C_\lambda(K;x,y)$ are different: the former involves the
squared conditional matrix expectation averaged over measurement
records, whereas the latter is a single average in the prior ensemble.

We now derive an exact identity for the measurement protocol defined
above. We consider a finite connected graph in the pure-gauge sector
and use the isotropic component measurements specified above.
We impose the matching condition
\begin{equation}
 K_{e,\rho}=\Delta_{e,\rho}
 \qquad\text{for each measured }(e,\rho).
 \label{eq:qd-HN-condition}
\end{equation}
Under this condition, we have
\begin{equation}
 \boxed{Q_\lambda(K,K;x,y)=C_\lambda(K;x,y)
 \qquad\forall\lambda.}
 \label{eq:qd-HN-identity}
\end{equation}
The proof proceeds in two steps. A group average cancels the posterior
normalization factor, after which the Gaussian integrals reduce the
result to the clean correlator.
We introduce the shifted Gaussian vectors
\begin{equation}
 \boldsymbol J_{e,\rho}
 =K_{e,\rho}\boldsymbol f_\rho(1_G)
  +\Delta_{e,\rho}\boldsymbol m_{e,\rho},
 \qquad
 S_J(\boldsymbol g)
 =\sum_{e=(i,j),\rho\in\mathcal S}
 \boldsymbol J_{e,\rho}\cdot\boldsymbol f_\rho(g_ig_j^{-1}).
 \label{eq:qd-shift}
\end{equation}
We also define the record-dependent partition function and the
unnormalized matrix correlator as
\begin{align}
 Z(J)&:=\int\!\mathcal Dg\,e^{S_J(\boldsymbol g)},\notag\\
 \mathcal N_\lambda(J)&:=\int\!\mathcal Dg\,
 e^{S_J(\boldsymbol g)}D^\lambda(g_xg_y^{-1}).
 \label{eq:qd-J-partition}
\end{align}
Then $A_\lambda(J)=\mathcal N_\lambda(J)/Z(J)$.

Let $d\nu_\Delta(J)$ denote the normalized product of centered, isotropic Gaussian
measures for the real vectors $\boldsymbol J_{e,\rho}$, each with covariance $\Delta_{e,\rho}I_{n_\rho}$.
Under the matching condition, completing the square in the Born
probability density gives
\begin{equation}
 p(J)\,dJ=\frac{e^{-\sum_{e,\rho}\Delta_{e,\rho}}}{Z_K}
  d\nu_\Delta(J)\,e^{S_J(\boldsymbol{1}_G)}Z(J),
 \label{eq:qd-HN-born}
\end{equation}
Here, $\boldsymbol{1}_G$ denotes the vertex configuration with $g_i=1_G$ at every site.
The factor $Z(J)$ comes from averaging over the underlying group
configuration. Substituting this probability density into the matrix
EA correlator leaves one inverse power of $Z(J)$:
\begin{equation}
 Q_\lambda=\frac{e^{-\sum_{e,\rho}\Delta_{e,\rho}}}{d_\lambda Z_K}
  \int d\nu_\Delta(J)\,e^{S_J(\boldsymbol{1}_G)}
  \frac{\operatorname{Tr}[\mathcal N_\lambda(J)
                    \mathcal N_\lambda(J)^\dagger]}{Z(J)}.
 \label{eq:qd-before-gauge-average}
\end{equation}

For each $\rho$, we combine the real and imaginary components of $\boldsymbol J_{ij,\rho}$
into a complex matrix $\mathsf J_{ij,\rho}$. We then make the simultaneous change of
variables $g_i\mapsto h_i g_i$ and $\mathsf J_{ij,\rho}\mapsto
D^\rho(h_i)\mathsf J_{ij,\rho}D^\rho(h_j)^\dagger$.
Viewed in real coordinates, this matrix transformation is orthogonal.
It leaves the centered isotropic Gaussian measure, $Z(J)$, and $\operatorname{Tr}[\mathcal N_\lambda(J)
\mathcal N_\lambda(J)^\dagger]$
unchanged. The factor $e^{S_J(\boldsymbol{1}_G)}$ is not invariant, but averaging it over
all $h_i\in G$ gives
\begin{equation}
 \int\!\mathcal Dh\,e^{S_{J^h}(\boldsymbol{1}_G)}
 =\int\!\mathcal Dh\,e^{S_J(\boldsymbol h^{-1})}
 =Z(J).
 \label{eq:qd-gauge-average}
\end{equation}
The last equality follows from the invariance of the normalized sum
over $\boldsymbol h$ under inversion. This factor cancels the remaining $1/Z(J)$
in~\eqref{eq:qd-before-gauge-average}.

Expanding $\mathcal N_\lambda(J)\mathcal N_\lambda(J)^\dagger$ introduces two independently summed vertex
configurations, $\boldsymbol g^{(1)}$ and $\boldsymbol g^{(2)}$. For each edge and representation,
the Gaussian integral together with~\eqref{eq:qd-feature-identity} gives
\begin{align}
 \int d\nu_{\Delta_{e,\rho}}(J)
 \exp\!\left\{\boldsymbol J\cdot
      [\boldsymbol f_\rho(u^{(1)})+\boldsymbol f_\rho(u^{(2)})]\right\}=\exp\!\left[\Delta_{e,\rho}
 \bigl\{1+c_\rho(u^{(1)}(u^{(2)})^{-1})\bigr\}\right].
 \label{eq:qd-two-copy-gaussian}
\end{align}
The factor $e^{\Delta_{e,\rho}}$ cancels the corresponding constant in the Born
probability density. We now introduce the relative vertex variables
\begin{equation}
 k_i=(g_i^{(2)})^{-1}g_i^{(1)}.
\end{equation}
For $u_{ij}^{(a)}=g_i^{(a)}(g_j^{(a)})^{-1}$, the product of the two edge variables can be written as
\begin{equation*}
 u_{ij}^{(1)}(u_{ij}^{(2)})^{-1}
 =g_i^{(2)}(k_i k_j^{-1})(g_i^{(2)})^{-1}.
\end{equation*}
The same relation holds at the endpoints, with $(i,j)$ replaced by $(x,y)$.
Since characters are invariant under conjugation, the interaction
obtained from the Gaussian integral depends only on $k_i k_j^{-1}$.
The matrix insertion similarly reduces to the normalized character
$\chi_\lambda(k_xk_y^{-1})/d_\lambda$. The change of variables $(\boldsymbol g^{(1)},\boldsymbol g^{(2)})\mapsto
(\boldsymbol k,\boldsymbol g^{(2)})$ is bijective, and the normalized
sum over $\boldsymbol g^{(2)}$ gives one. The remaining sum over~$\boldsymbol k$ is the clean
correlator in~\eqref{eq:qd-clean-correlator}, completing the proof of~\eqref{eq:qd-HN-identity}.

\subsection{Hierarchy of EA correlators}
\label{qd:sec:hierarchy}
In this subsection, we study the clean correlator $C_{\lambda}(K;x,y)$ and derive a
hierarchy among the EA correlators of different electric charges.
To this end, we use the Fortuin--Kasteleyn--Ginibre (FKG) inequality~\cite{Fortuin1971Correlation}
and expand the interactions in terms of the normal subgroups of
$G$, denoted by $N$.
Recalling from~\eqref{eq:qd-prior} that the clean probability distribution is determined
by the edge interactions \(V_e(u):=\sum_{\rho\in S}K_{e,\rho}c_\rho(u)\), where \(u=g_i g_j^{-1}\) for an oriented edge \(e=(i,j)\), 
we impose an additional condition on these interactions: each \(V_e(u)\)
is assumed to admit an expansion in the so-called indicator functions of normal
subgroups of $G$, $\delta_N(u)$
with nonnegative coefficients, up to an additive
constant $c_e$. Namely, 
\begin{equation}
V_e(u):=\sum_{\rho\in\mathcal S}K_{e,\rho}c_\rho(u)
=c_e+\sum_{N\triangleleft G}J_{e,N}\,\delta_N(u),
\qquad J_{e,N}\geq0,
\label{qd:eq:positive}
\end{equation}
where $\delta_N(u)=1$ for $u\in N$ and $0$ otherwise.
We work on the higher Nishimori manifold $K_{e,\rho}=\Delta_{e,\rho}$ and assume the
nonnegative expansion~\eqref{qd:eq:positive}. \par

Let $\mathcal J$ be the finite set of pairs
$\alpha=(e,N)$ for which $J_{e,N}$ is strictly positive. Here, $e=(i,j)$ is an oriented
edge and $N\triangleleft G$ is a normal subgroup. Terms with $J_{e,N}=0$ can be omitted.
For each pair, we define $J_\alpha:=J_{e,N}$ and $\delta_\alpha(\boldsymbol g):=\delta_N(g_i g_j^{-1})$.
Thus, $\delta_\alpha(\boldsymbol g)=1$ when the group variables at the endpoints of~$e$
represent the same element of the quotient group $G/N$.

Since $\delta_\alpha$ takes only the values zero and one, we can expand its
Boltzmann factor as
\begin{equation}\label{eq:qdC-212}
 e^{J_\alpha\delta_\alpha(\boldsymbol g)}
 =1+v_\alpha\delta_\alpha(\boldsymbol g),
 \qquad v_\alpha:=e^{J_\alpha}-1\geq0.
\end{equation}
Both sides are $1$ for $\delta_\alpha=0$ and $e^{J_\alpha}$ for $\delta_\alpha=1$.
The terms $c_e$ in~\eqref{qd:eq:positive} are independent of the configuration and
contribute only an overall factor. Omitting this factor, we write the
Boltzmann weight as
\begin{equation}\label{eq:qdC-213}
 W(\boldsymbol g)
 :=\prod_{\alpha\in\mathcal J}
 \bigl[1+v_\alpha\delta_\alpha(\boldsymbol g)\bigr].
\end{equation}
For each factor, we choose either $1$ or $v_\alpha\delta_\alpha$.
Let $B\subseteq\mathcal J$ be the set of labels for which the second term is chosen.
The product in~\eqref{eq:qdC-213} then becomes
\begin{equation*}
 W(\boldsymbol g)
 =\sum_{B\subseteq\mathcal J}
 \left(\prod_{\alpha\in B}v_\alpha\right)
 \left(\prod_{\alpha\in B}\delta_\alpha(\boldsymbol g)\right).
\end{equation*}
We define
\begin{equation}\label{eq:qdC-214}
 H_B:=\bigl\{\boldsymbol g\in G^V:
      \delta_\alpha(\boldsymbol g)=1
      \text{ for every }\alpha\in B\bigr\}.
\end{equation}
Summing over the group spins, we obtain
\begin{equation*}
 \mathcal Z
 :=\sum_{\boldsymbol g\in G^V}W(\boldsymbol g)
 =\sum_{B\subseteq\mathcal J}
    |H_B|\prod_{\alpha\in B}v_\alpha.
\end{equation*}
The clean partition function in~\eqref{eq:qd-deformed-state} is therefore $Z_K=e^{\sum_e c_e}|G|^{-|V|}\mathcal Z$.
The terms in this expansion have nonnegative weights, which we
normalize to define the probability distribution
\begin{equation}\label{eq:qdC-215}
 \mu(B):=
 \frac{|H_B|\displaystyle\prod_{\alpha\in B}v_\alpha}
 {\displaystyle\sum_{D\subseteq\mathcal J}
  |H_D|\prod_{\alpha\in D}v_\alpha},
 \qquad \sum_{B\subseteq\mathcal J}\mu(B)=1.
\end{equation}
Applying the same expansion to the numerator of the clean correlator
in~\eqref{eq:qd-clean-correlator}, we obtain
\begin{equation}\label{eq:qdC-216}
 \begin{aligned}
 C_\lambda(K;x,y)
=\sum_{B\subseteq\mathcal J}\mu(B)
   C_\lambda^{(B)}(x,y)
\end{aligned}
\end{equation}
with 
\begin{eqnarray}
     C_\lambda^{(B)}(x,y)
:=\frac{1}{|H_B|}\sum_{\boldsymbol g\in H_B}
     \frac{\chi_\lambda(g_xg_y^{-1})}{d_\lambda}.\label{CBX}
\end{eqnarray}

To evaluate the correlator in~\eqref{eq:qdC-216}, we introduce the following set
for fixed $B$ and endpoints $x,y$:
\begin{eqnarray}
    N_B(x,y)
 :=\bigl\{g_xg_y^{-1}:\boldsymbol g\in H_B\bigr\}.\label{nbb}
\end{eqnarray}
Here $N_B(x,y)$ is the set of relative group elements allowed by
the constraints in $B$. We omit the endpoints when they are clear.
In Appendix~\ref{nb}, we show that $N_B(x,y)$ is a normal subgroup of $G$ and that a uniform sample from $H_B$ induces the uniform distribution
on $N_B(x,y)$. The constrained character average is therefore
\begin{equation}\label{eq:qdC-217}
 \begin{aligned}
 C_\lambda^{(B)}(x,y)
 =\frac{1}{d_\lambda|N_B|}
   \sum_{n\in N_B}\chi_\lambda(n)=\begin{cases}
      1,&N_B\subseteq\ker D^\lambda,\\
      0,&\text{otherwise}.
    \end{cases}
 \end{aligned}
\end{equation}
We further define the event 
\begin{equation*}
 \mathcal A_\lambda(x,y):=
 \bigl\{B\subseteq\mathcal J:N_B(x,y)\subseteq\ker D^\lambda\bigr\}.
\end{equation*}
Let $\mathbf 1_{\mathcal A_\lambda}(B)$ equal one when
$B\in\mathcal A_\lambda$ and zero otherwise.
Combining the constrained average with $Q_\lambda=C_\lambda$, we obtain
%
\begin{equation}\label{eq:qdC-219}
 Q_\lambda(x,y)=C_\lambda(x,y)
 =\sum_{B\subseteq\mathcal J}
   \mu(B)\mathbf1_{\mathcal A_\lambda}(B)
 =\mu\bigl(\mathcal A_\lambda\bigr).
\end{equation}
This expresses the EA correlator as the probability of an event in
the constraint ensemble.

We can now compare the correlators of two representations.
If $\ker D^\lambda\subseteq\ker D^\nu$, then $\mathcal A_\lambda\subseteq\mathcal A_\nu$. The inclusion of these events gives
\begin{equation}\label{eq:qdC-220}
 \ker D^\lambda\subseteq\ker D^\nu
 \quad\Longrightarrow\quad Q_\lambda(x,y)\leq Q_\nu(x,y).
\end{equation}
Next, let $\nu\prec\lambda\otimes\eta$, meaning that $\nu$
appears in the tensor-product decomposition of $\lambda$ and $\eta$.
If every element of $N_B$ acts trivially in both representations,
it also acts trivially in each constituent of their tensor product.
Hence $\mathcal A_\lambda\cap\mathcal A_\eta\subseteq
\mathcal A_\nu$.
Adding constraints can only shrink $N_B$, so an event
$\mathcal A_\lambda$ that already holds continues to hold.
The events $\mathcal A_\lambda$ and $\mathcal A_\eta$ are
therefore increasing under inclusion of constraint sets.
As discussed in Appendix~\ref{FKG}, one can make use of  the FKG inequality, which gives
\begin{eqnarray}
     \mu(\mathcal A_\lambda\cap\mathcal A_\eta)
 \geq \mu(\mathcal A_\lambda)\mu(\mathcal A_\eta).
\end{eqnarray}
Combining this inequality with the event inclusion and the probability
representation of the correlators yields
%
\begin{equation}\label{eq:qdC-221}
 \nu\prec\lambda\otimes\eta
 \quad\Longrightarrow\quad
 Q_\nu(x,y)\geq Q_\lambda(x,y)Q_\eta(x,y).
\end{equation}
The relations~\eqref{eq:qdC-220} and~\eqref{eq:qdC-221} give the hierarchy among the EA
correlators of different electric charges. We illustrate these bounds in the quantum doubles of $S_3$ and $D_4$ below. 
\subsection{Examples}
We now apply the above inequalities to two concrete examples, $S_3$ and $D_4$ quantum double.
\subsubsection{$G=S_3$}
The group $S_3$ consists of the six permutations of three objects.
Choosing $r=(123)$ and $s=(12)$, we write
\begin{equation*}
 S_3=\langle r,s:r^3=s^2=1_G,\ srs=r^{-1}\rangle
     =\{1_G,r,r^2,s,rs,r^2s\}.
\end{equation*}
The even permutations form a subgroup
\begin{equation*}
 A_3:=\{1_G,r,r^2\}=\langle r\rangle\cong\mathbb Z_3.
\end{equation*}
This subgroup is normal: conjugation by $r$ preserves it, while
conjugation by $s$ exchanges $r$ and $r^{-1}=r^2$.

The three irreducible representations of $S_3$ label its three pure
electric sectors, including the vacuum.
We denote the trivial, sign, and standard two dimensional
representations by $A$, $B$, and $C$, respectively.
Their dimensions are $d_A=d_B=1$ and $d_C=2$. The sign representation
satisfies $D^B(r)=1$ and $D^B(s)=-1$.
The characters and conjugacy class sizes are listed in
Table~\ref{tab:v4-s3-characters}.
\begin{table}[htbp]
\centering
\begin{tabular}{c|ccc}
 & $\{1_G\}$ & $\{r,r^2\}$ & $\{s,rs,r^2s\}$ \\
\hline
class size & 1 & 2 & 3 \\
$\chi_A$ & 1 & 1 & 1 \\
$\chi_B$ & 1 & 1 & $-1$ \\
$\chi_C$ & 2 & $-1$ & 0
\end{tabular}
\caption{Character table of $S_3$, including sizes of the conjugacy classes.
The representations $A$, $B$, and $C$ are the trivial, sign,
and two dimensional irreducible representations, respectively.}
\label{tab:v4-s3-characters}
\end{table}\par
Characters multiply under tensor products and add under direct sums.
The electric fusion rules are therefore
\begin{equation}\label{eq:qd9-s3-fusion}
\begin{gathered}
 A\otimes\rho=\rho\quad(\rho\in\{A,B,C\}),\qquad
 B\otimes B=A,\qquad B\otimes C=C,\\
 C\otimes C=A\oplus B\oplus C.
\end{gathered}
\end{equation}

In this quantum double, we think of weak measurement of two electric charges, 
corresponding to two representations $\mathcal S=\{B,C\}$. 
Since their characters are real, the corresponding classical spin  interaction in~\eqref{eq:qd-c-and-norm} gives
\begin{equation}
 V(u)=K_B\chi_B(u)+\frac{K_C}{2}\chi_C(u).\label{vu1}
\end{equation}
To apply the hierarchy, we express this interaction in terms of
indicator functions of normal subgroups. The two nonconstant indicators needed here
are $\delta_{A_3}$ and $\delta_{\{1_G\}}$, whose values are listed in
Table~\ref{tab:v4-s3-indicators}.
\begin{table}[htbp]
\centering
\begin{tabular}{c|cc}
$u$ & $\delta_{A_3}(u)$ & $\delta_{\{1_G\}}(u)$ \\
\hline
$1_G$ & 1 & 1 \\
$r,r^2$ & 1 & 0 \\
$s,rs,r^2s$ & 0 & 0
\end{tabular}
\caption{Indicator functions for the proper normal subgroups
$A_3$ and $\{1_G\}$ of $S_3$. Their values are constant
on each conjugacy class.}
\label{tab:v4-s3-indicators}
\end{table}\par
Comparing Tables~\ref{tab:v4-s3-characters} and
\ref{tab:v4-s3-indicators}, we obtain
\begin{equation*}
 \chi_B(u)=2\delta_{A_3}(u)-1,\qquad
 \chi_C(u)=3\delta_{\{1_G\}}(u)-\delta_{A_3}(u).
\end{equation*}
Substituting these identities into~\eqref{vu1}, we obtain
\begin{equation}
\begin{aligned}
 V(u)
 =K_B\bigl[2\delta_{A_3}(u)-1\bigr]
   +\frac{K_C}{2}\bigl[3\delta_{\{1_G\}}(u)-\delta_{A_3}(u)\bigr]
 =-K_B+\left(2K_B-\frac{K_C}{2}\right)\delta_{A_3}(u)
        +\frac{3K_C}{2}\delta_{\{1_G\}}(u).\label{eq:qd9-s3-expansion}
\end{aligned}
\end{equation}
To apply the hierarchy derived
in the preceding subsection, we require the coefficients in this
expansion to be nonnegative:
\begin{equation*}
 K_C\geq0,\qquad K_B\geq K_C/4.
\end{equation*}
On the higher Nishimori manifold, these conditions become
\begin{equation*}
 K_B=\Delta_B,\qquad K_C=\Delta_C,\qquad
 \Delta_C\geq0,\qquad\Delta_B\geq\Delta_C/4.
\end{equation*}
From~\eqref{eq:qdC-220} and $\ker D^C=\{1_G\}\subseteq A_3=\ker D^B$, we have
\begin{eqnarray}
     Q_B(x,y)\geq Q_C(x,y).
\end{eqnarray}
Thus, in this parameter region, long range learnability of the
non-Abelian electric charge $C$ implies long range learnability
of the Abelian electric charge $B$.
\subsubsection{$G=D_4$}
We next consider $D_4$, given by
\begin{equation*}
 D_4=\langle r,s:r^4=s^2=1_G,\ srs=r^{-1}\rangle
     =\{r^m s^n:0\leq m\leq3,\ n=0,1\}.
\end{equation*}
Here, $r$ is a quarter-turn and $s$ is a reflection of a square.
Besides the trivial representation $\mathbf1$, there are three
one-dimensional representations $a,b,c=ab$, defined by
\begin{equation*}
 a(r^m s^n)=(-1)^m,\qquad b(r^m s^n)=(-1)^n,\qquad
 c(r^m s^n)=(-1)^{m+n}.
\end{equation*}
The two dimensional representation $E$ can be chosen as
\begin{equation*}
 D^E(r)=\begin{pmatrix}0&-1\\1&0\end{pmatrix},\qquad
 D^E(s)=\begin{pmatrix}1&0\\0&-1\end{pmatrix}.
\end{equation*}
The five irreducible representations $1,a,b,c,E$ label the pure
electric sectors of $D(D_4)$. The trivial representation is the vacuum,
while $a,b,c,E$ label the four nontrivial electric charges.

The characters are listed in Table~\ref{tab:v4-d4-characters}.
\begin{table}[htbp]
\centering
\begin{tabular}{c|ccccc}
 & $\{1_G\}$ & $\{r^2\}$ & $\{r,r^3\}$
 & $\{s,r^2s\}$ & $\{rs,r^3s\}$ \\
\hline
class size & 1 & 1 & 2 & 2 & 2 \\
$\chi_1$ & 1 & 1 & 1 & 1 & 1 \\
$\chi_a$ & 1 & 1 & $-1$ & 1 & $-1$ \\
$\chi_b$ & 1 & 1 & 1 & $-1$ & $-1$ \\
$\chi_c$ & 1 & 1 & $-1$ & $-1$ & 1 \\
$\chi_E$ & 2 & $-2$ & 0 & 0 & 0
\end{tabular}
\caption{Character table of $D_4$, the dihedral group of order eight,
including sizes of the conjugacy classes. The representations $1,a,b,c$
are one-dimensional, while $E$ is two dimensional.}
\label{tab:v4-d4-characters}
\end{table}\par

Since the one-dimensional characters take values $\pm1$, multiplying
their rows gives
\begin{equation}\label{eq:qd9-d4-abelianfusion}
 a\otimes a=b\otimes b=c\otimes c=\mathbf1,\qquad
 a\otimes b=c,\quad b\otimes c=a,\quad c\otimes a=b.
\end{equation}
Each of these characters is $+1$ at $1_G$ and $r^2$, the only
classes on which $\chi_E$ is nonzero. Hence,
\begin{equation}\label{eq:qd9-d4-mixedfusion}
 \alpha\otimes E=E\qquad(\alpha\in\{\mathbf1,a,b,c\}).
\end{equation}
We also have 
\begin{equation}\label{eq:qd9-d4-efusion}
 E\otimes E=\mathbf1\oplus a\oplus b\oplus c.
\end{equation}
The kernels of these representations are
\begin{equation*}
\begin{aligned}
 \ker D^a&=\{1_G,r^2,s,r^2s\},&
 \ker D^b&=\{1_G,r,r^2,r^3\},\\
 \ker D^c&=\{1_G,r^2,rs,r^3s\},&
 \ker D^E&=\{1_G\}.
\end{aligned}
\end{equation*}
The three kernels of order four are normal, and none is contained
in another. Their intersection is the center $Z(D_4)=\{1_G,r^2\}$.

In the learning problem of the $D_4$ quantum double, we weakly measure four electric charges, 
 $\mathcal S=\{a,b,c,E\}$. The  corresponding classical spin interaction~\eqref{eq:qd-c-and-norm} reads
\begin{eqnarray}
     V(u)=K_a a(u)+K_b b(u)+K_c c(u)+\frac{K_E}{2}\chi_E(u).\label{hugo}
\end{eqnarray}
We again express the interaction in terms of indicator functions of normal subgroups.
The five nonconstant indicators are
$\delta_{\ker D^\alpha}$ for $\alpha=a,b,c$,
$\delta_{Z(D_4)}$, and $\delta_{\{1_G\}}$.
For a one-dimensional character $\chi_\alpha\;(\alpha=a,b,c)$, it is straightforward to verify that 
\begin{equation*}
 \alpha(u)=2\delta_{\ker D^\alpha}(u)-1.
\end{equation*}
Also, one can verify that 
\begin{equation*}
 \frac{\chi_E(u)}{2}=2\delta_{\{1_G\}}(u)-\delta_{Z(D_4)}(u),\qquad
 \delta_{Z(D_4)}(u)=\frac{1+a(u)+b(u)+c(u)}{4}.
\end{equation*}
Substituting these into~\eqref{hugo}, we have
\begin{equation*}
 V(u)=2K_E\delta_{\{1_G\}}(u)-\frac{K_E}{4}
      +\sum_{\alpha=a,b,c}\left(K_\alpha-\frac{K_E}{4}\right)\alpha(u).
\end{equation*}
which is further rewritten as
\begin{equation}\label{eq:qd9-d4-expansion}
 V(u)=c_0+2K_E\delta_{\{1_G\}}(u)
       +\sum_{\alpha=a,b,c}\left(2K_\alpha-\frac{K_E}{2}\right)
           \delta_{\ker D^\alpha}(u),
\end{equation}
with $c_0=-(K_a+K_b+K_c)+K_E/2$. \par

To apply the hierarchy relation regarding the EA correlators of electric charges, we require
\begin{equation}
 K_E\geq0,\qquad K_a,K_b,K_c\geq K_E/4.
\end{equation}
On the higher Nishimori manifold, these conditions read $\Delta_E\geq0$ and
$\Delta_a,\Delta_b,\Delta_c\geq\Delta_E/4$. In this parameter region, the relation~\eqref{eq:qdC-220} yields
\begin{equation}
 Q_a(x,y)\geq Q_E(x,y),\qquad Q_b(x,y)\geq Q_E(x,y),\qquad Q_c(x,y)\geq Q_E(x,y).
\end{equation}
Also, from~\eqref{eq:qdC-221}, one finds
\begin{equation}
 Q_c(x,y)\geq Q_a(x,y)Q_b(x,y),\qquad Q_a(x,y)\geq Q_b(x,y)Q_c(x,y),\qquad Q_b(x,y)\geq Q_c(x,y)Q_a(x,y).
\end{equation}
In this parameter region, long range learnability of the non-Abelian
electric charge $E$ implies that of all three nontrivial Abelian
electric charges. The fusion bounds further imply that learnability
of any two of~$a,b,c$ entails learnability of the third.

\section{Outlook}\label{oput}
In this work, we formulated Bayesian learning under local weak
measurements for a broad class of CSS stabilizer codes and extended the
analysis to finite group quantum double models.
For CSS codes, matching the wavefunction deformation and Gaussian
measurement strengths on the higher Nishimori manifold leads to an
enlarged $S_{R+1}$ replica permutation symmetry and a local gauge
redundancy.
These structures yield exact identities among record averaged
correlators that quantify the learnability of nonlocal observables
associated with fractionalized excitations.
In particular, these identities relate the learning correlators to
correlations in the corresponding clean models.
For nonnegative ferromagnetic couplings, classical correlation
inequalities then establish exact hierarchies of learnability:
simultaneous long range learnability of two observables implies that
of their product.

The rank-2 toric code illustrates this mechanism through spatially
modulated symmetries.
Acting a lattice translation on a dipole loop operator gives a homogeneous monopole loop
operator.
Applying the hierarchy to the corresponding open string observables
shows that long range dipole learnability implies long range
learnability of both associated monopole observables.
When the relevant correlators exhibit critical power law scaling,
the same inequalities constrain their scaling dimensions.
In the X-cube model discussed in Appendix~\ref{xcube}, the same approach
applies to planar subsystem symmetries.
Multiplication relations among observables invariant under the subsystem symmetry yield
a hierarchy under spatial subdivision and cyclic inequalities among
extended observables formed from products of open strings.

For quantum double models, we established an analogous
higher Nishimori identity for pure electric observables.
When the classical interactions can each be written as a constant
plus a sum of indicator functions of normal subgroups with
nonnegative coefficients, representation kernels and the fusion
rules of anyons with electric charges yield further inequalities among
learning correlators.
These inequalities apply to both Abelian and non-Abelian electric
charges.
In the parameter regions considered for the $S_3$ and $D_4$
examples, learning the non-Abelian electric anyons implies learning the associated nontrivial Abelian
anyons.
Together with the rank-2 toric code results, these examples show
how learnability hierarchies reflect both the spatial structure
of observables and the electric charges they probe.
Operator multiplication and fusion thus constrain which types
of nonlocal information can be learned simultaneously, providing
a guide to learning phase diagrams and critical behavior in
quantum codes.

Several questions remain open.
One direction is to determine which aspects of the hierarchy survive
away from the exactly matched higher Nishimori manifold and under more
general measurement protocols.
In Appendix~\ref{stability}, we briefly argue that  the decoupled critical point
of the rank-2 toric code is unstable under plaquette deformation
and measurement.
The hierarchy remains exact when the higher Nishimori condition
is preserved, but its persistence away from this condition
remains an open question.
In particular, weaker constraints on simultaneous long range
learnability may survive even when the exact correlator inequalities
are no longer guaranteed.
A second direction is to extend the quantum double analysis beyond
pure electric observables to observables associated with magnetic and
dyonic excitations.
More broadly, it would be interesting to investigate analogous learning
problems in topological models with categorical or higher group
symmetries, and to determine whether their operator and fusion
structures give rise to corresponding identities and learnability
hierarchies. 
\section*{Acknowledgments}
We thank W. Cao, J. Haruna, B. Han, K. Kato, R. A. Patil, S. Trebst,
H. Yan, G. Y. Zhu for helpful discussions. We also thank Bo Han for comments on the draft. 
This work is in part supported by  JST CREST (Grant No. JPMJCR24I3), JSPS KAKENHI Grant Number JP22H01222 and the Royal Society grants ICA/R2/242058 and IEC/R3/243103.
We are also grateful for the valuable discussions and feedback from the participants of the UK-Japan Workshop on Quantum Gravity 2026 at Durham University.

\bibliography{main}

@article{Vadali:2026,
    author = "Vadali, Avi and Vanhove, Robijn and Verresen, Ruben and Alicea, Jason and Sala, Pablo",
    title = "{Statistical Mechanics and Symmetries of Non-Abelian Anyon Proliferation: From Deformation to Decoherence}",
    eprint = "2606.12527",
    archivePrefix = "arXiv",
    primaryClass = "quant-ph",
    month = "6",
    year = "2026"
}

@article{Cui:2019lvb,
    author = "Cui, Shawn X. and Ding, Dawei and Han, Xizhi and Penington, Geoffrey and Ranard, Daniel and Rayhaun, Brandon C. and Shangnan, Zhou",
    title = "{Kitaev's quantum double model as an error correcting code}",
    eprint = "1908.02829",
    archivePrefix = "arXiv",
    primaryClass = "quant-ph",
    doi = "10.22331/q-2020-09-24-331",
    journal = "Quantum",
    volume = "4",
    pages = "331",
    year = "2020"
}

@inproceedings{Wootton:2010dah,
    author = "Wootton, James R. and Lahtinen, Ville and Pachos, Jiannis K.",
    title = "{Universal quantum computation with a non-Abelian topological memory}",
    eprint = "0906.2748",
    archivePrefix = "arXiv",
    primaryClass = "quant-ph",
    month = "10",
    year = "2010"
}

@article{Fortuin1971Correlation,
  author    = {Fortuin, C. M. and Kasteleyn, P. W. and Ginibre, J.},
  title     = {Correlation inequalities on some partially ordered sets},
  journal   = {Communications in Mathematical Physics},
  volume    = {22},
  number    = {2},
  pages     = {89--103},
  year      = {1971},
  doi       = {10.1007/BF01651330},
  url       = {https://link.springer.com/article/10.1007/BF01651330}
}

@article{bravyi2022adaptive,
  title={Adaptive constant-depth circuits for manipulating non-Abelian anyons},
  author={Bravyi, Sergey and Kim, Isaac and Kliesch, Alexander and Koenig, Robert},
  journal={arXiv preprint arXiv:2205.01933},
  year={2022}
}

@article{jing2026integer,
  title={Integer Linear Programming Decoder for Abelian and Non-Abelian Topological Codes},
  author={Jing, Dian and Zhang, Aubrey and Jiang, Liang and Verresen, Ruben},
  journal={arXiv preprint arXiv:2608.18512},
  year={2026}
}

@article{kim2026transparent,
  title={Transparent Domain Walls through Information Convex Sets},
  author={Kim, Jintae and Lamas, Amanda Gatto and Gliozzi, Jacopo and Shi, Bowen and Hughes, Taylor L and Lee, Jong Yeon},
  journal={arXiv preprint arXiv:2609.26892},
  year={2026}
}

@article{PhysRevB.110.045146,
  title = {Entanglement entropy of topological phases with multipole symmetry},
  author = {Ebisu, Hiromi},
  journal = {Phys. Rev. B},
  volume = {110},
  issue = {4},
  pages = {045146},
  numpages = {24},
  year = {2024},
  month = {Jul},
  publisher = {American Physical Society},
  doi = {10.1103/PhysRevB.110.045146},
  url = {https://link.aps.org/doi/10.1103/PhysRevB.110.045146}
}

@article{sala2026climbing,
  title={Climbing the Fusion Tree: Center Ordering and Decoding Hierarchy in Non-Abelian Mixed States},
  author={Sala, Pablo and Temkin, Vlad and Xu, Cenke and Podolsky, Daniel and Altman, Ehud},
  journal={arXiv preprint arXiv:2610.06692},
  year={2026}
}

@article{7dpt-d4s5,
  title = {Bayesian critical points in classical lattice models},
  author = {Nahum, Adam and Jacobsen, Jesper Lykke},
  journal = {Phys. Rev. B},
  volume = {112},
  issue = {23},
  pages = {235113},
  numpages = {57},
  year = {2025},
  month = {Dec},
  publisher = {American Physical Society},
  doi = {10.1103/7dpt-d4s5},
  url = {https://link.aps.org/doi/10.1103/7dpt-d4s5}
}

@article{PhysRevB.111.035149,
  title = {Noninvertible operators in one, two, and three dimensions via gauging spatially modulated symmetry},
  author = {Ebisu, Hiromi and Han, Bo},
  journal = {Phys. Rev. B},
  volume = {111},
  issue = {3},
  pages = {035149},
  numpages = {24},
  year = {2025},
  month = {Jan},
  publisher = {American Physical Society},
  doi = {10.1103/PhysRevB.111.035149},
  url = {https://link.aps.org/doi/10.1103/PhysRevB.111.035149}
}

@article{PhysRevB.78.115421,
  title = {Family of non-Abelian Kitaev models on a lattice: Topological condensation and confinement},
  author = {Bombin, H. and Martin-Delgado, M. A.},
  journal = {Phys. Rev. B},
  volume = {78},
  issue = {11},
  pages = {115421},
  numpages = {28},
  year = {2008},
  month = {Sep},
  publisher = {American Physical Society},
  doi = {10.1103/PhysRevB.78.115421},
  url = {https://link.aps.org/doi/10.1103/PhysRevB.78.115421}
}

@article{KovalevPryadko2013,
  author  = {Kovalev, Alexey A. and Pryadko, Leonid P.},
  title   = {Spin glass reflection of the decoding transition for quantum error correcting codes},
  journal = {arXiv},
  year    = {2013},
  doi     = {10.48550/arxiv.1311.7688}
}

@article{chubb2021statistical,
  author   = {Chubb, Christopher T. and Flammia, Steven T.},
  title    = {Statistical mechanical models for quantum codes with correlated noise},
  journal  = {Annales de l'Institut Henri Poincar{\'e} D},
  volume   = {8},
  number   = {2},
  pages    = {253--305},
  year     = {2021},
  doi      = {10.4171/AIHPD/105},
  url      = {https://ems.press/journals/aihpd/articles/1146536}
}

@article{PhysRevLett.103.090501,
  title = {Error Threshold for Color Codes and Random Three-Body Ising Models},
  author = {Katzgraber, Helmut G. and Bombin, H. and Martin-Delgado, M. A.},
  journal = {Phys. Rev. Lett.},
  volume = {103},
  issue = {9},
  pages = {090501},
  numpages = {4},
  year = {2009},
  month = {Aug},
  publisher = {American Physical Society},
  doi = {10.1103/PhysRevLett.103.090501},
  url = {https://link.aps.org/doi/10.1103/PhysRevLett.103.090501}
}

@article{Nishimori_1980,author   = {Nishimori, Hidetoshi},title    = {Exact results and critical properties of the Ising model with competing interactions},journal  = {Journal of Physics C: Solid State Physics},volume   = {13},number   = {21},pages    = {4071--4081},year     = {1980},month    = {jul},publisher = {IOP Publishing},doi      = {10.1088/0022-3719/13/21/012}}

@article{PhysRevB.40.9249,
  title = {\ensuremath{\epsilon} expansion for the Nishimori multicritical point of spin glasses},
  author = {Le Doussal, Pierre and Harris, A. Brooks},
  journal = {Phys. Rev. B},
  volume = {40},
  issue = {13},
  pages = {9249--9252},
  numpages = {0},
  year = {1989},
  month = {Nov},
  publisher = {American Physical Society},
  doi = {10.1103/PhysRevB.40.9249},
  url = {https://link.aps.org/doi/10.1103/PhysRevB.40.9249}
}

@article{PhysRevLett.61.625,
  title = {Location of the Ising Spin-Glass Multicritical Point on Nishimori's Line},
  author = {Le Doussal, Pierre and Harris, A. Brooks},
  journal = {Phys. Rev. Lett.},
  volume = {61},
  issue = {5},
  pages = {625--628},
  numpages = {0},
  year = {1988},
  month = {Aug},
  publisher = {American Physical Society},
  doi = {10.1103/PhysRevLett.61.625},
  url = {https://link.aps.org/doi/10.1103/PhysRevLett.61.625}
}

@article{iba1999nishimori,
  title={The Nishimori line and Bayesian statistics},
  author={Iba, Yukito},
  journal={Journal of Physics A: Mathematical and General},
  volume={32},
  number={21},
  pages={3875--3888},
  year={1999}
}

@article{PhysRevB.77.054433,
  title = {Quantum topological phase transition at the microscopic level},
  author = {Castelnovo, Claudio and Chamon, Claudio},
  journal = {Phys. Rev. B},
  volume = {77},
  issue = {5},
  pages = {054433},
  numpages = {14},
  year = {2008},
  month = {Feb},
  publisher = {American Physical Society},
  doi = {10.1103/PhysRevB.77.054433},
  url = {https://link.aps.org/doi/10.1103/PhysRevB.77.054433}
}

@article{johnston2017plaquette,
  title={Plaquette Ising models, degeneracy and scaling},
  author={Johnston, Desmond A and Mueller, Marco and Janke, Wolfhard},
  journal={The European Physical Journal Special Topics},
  volume={226},
  number={4},
  pages={749--764},
  year={2017},
  publisher={Springer}
}

@article{PhysRevLett.129.230502,
  title = {Optimal Thresholds for Fracton Codes and Random Spin Models with Subsystem Symmetry},
  author = {Song, Hao and Sch\"onmeier-Kromer, Janik and Liu, Ke and Viyuela, Oscar and Pollet, Lode and Martin-Delgado, M. A.},
  journal = {Phys. Rev. Lett.},
  volume = {129},
  issue = {23},
  pages = {230502},
  numpages = {7},
  year = {2022},
  month = {Nov},
  publisher = {American Physical Society},
  doi = {10.1103/PhysRevLett.129.230502},
  url = {https://link.aps.org/doi/10.1103/PhysRevLett.129.230502}
}

@article{5ywn-6d3q,
  title = {Decoherence and Wave-Function Deformation of ${D}_{4}$ Non-Abelian Topological Order},
  author = {Sala, Pablo and Alicea, Jason and Verresen, Ruben},
  journal = {Phys. Rev. X},
  volume = {15},
  issue = {3},
  pages = {031002},
  numpages = {43},
  year = {2025},
  month = {Jul},
  publisher = {American Physical Society},
  doi = {10.1103/5ywn-6d3q},
  url = {https://link.aps.org/doi/10.1103/5ywn-6d3q}
}

@article{PhysRevLett.129.120604,
  title = {Field Theory of Charge Sharpening in Symmetric Monitored Quantum Circuits},
  author = {Barratt, Fergus and Agrawal, Utkarsh and Gopalakrishnan, Sarang and Huse, David A. and Vasseur, Romain and Potter, Andrew C.},
  journal = {Phys. Rev. Lett.},
  volume = {129},
  issue = {12},
  pages = {120604},
  numpages = {7},
  year = {2022},
  month = {Sep},
  publisher = {American Physical Society},
  doi = {10.1103/PhysRevLett.129.120604},
  url = {https://link.aps.org/doi/10.1103/PhysRevLett.129.120604}
}

@article{dennis2002topological,
  title={Topological quantum memory},
  author={Dennis, Eric and Kitaev, Alexei and Landahl, Andrew and Preskill, John},
  journal={Journal of Mathematical Physics},
  volume={43},
  number={9},
  pages={4452--4505},
  year={2002},
  publisher={American Institute of Physics}
}

@article{PhysRevB.110.085158,
  title = {Tapestry of dualities in decohered quantum error correction codes},
  author = {Su, Kaixiang and Yang, Zhou and Jian, Chao-Ming},
  journal = {Phys. Rev. B},
  volume = {110},
  issue = {8},
  pages = {085158},
  numpages = {30},
  year = {2024},
  month = {Aug},
  publisher = {American Physical Society},
  doi = {10.1103/PhysRevB.110.085158},
  url = {https://link.aps.org/doi/10.1103/PhysRevB.110.085158}
}

@book{nishimori2001statistical,
  title={Statistical physics of spin glasses and information processing},
  author={Nishimori, Hidetoshi},
  volume={187},
  year={2001},
  publisher={Oxford university press Oxford}
}

@article{zdeborova2016statistical,
  title={Statistical physics of inference: Thresholds and algorithms},
  author={Zdeborov{\'a}, Lenka and Krzakala, Florent},
  journal={Advances in Physics},
  volume={65},
  number={5},
  pages={453--552},
  year={2016},
  publisher={Taylor \& Francis}
}

@article{KITAEV20032,
title = {Fault-tolerant quantum computation by anyons},
journal = {Annals of Physics},
volume = {303},
number = {1},
pages = {2-30},
year = {2003},
issn = {0003-4916},
doi = {https://doi.org/10.1016/S0003-4916(02)00018-0},
url = {https://www.sciencedirect.com/science/article/pii/S0003491602000180},
author = {A.Yu. Kitaev}
}

@article{Lyons:2024mapping,
  title={Understanding stabilizer codes under local decoherence through a general statistical mechanics mapping},
  author={Lyons, Anasuya},
  journal={arXiv preprint arXiv:2403.03955},
  year={2024}
}

@article{Putz:2026learning,
  title = {Learning Transitions in Classical Ising Models and Deformed Toric Codes},
  author = {P\"utz, Malte and Garratt, Samuel J. and Nishimori, Hidetoshi and Trebst, Simon and Zhu, Guo-Yi},
  journal = {Phys. Rev. Lett.},
  volume = {136},
  issue = {19},
  pages = {190402},
  numpages = {9},
  year = {2026},
  month = {May},
  publisher = {American Physical Society},
  doi = {10.1103/4dwm-kn11},
  url = {https://link.aps.org/doi/10.1103/4dwm-kn11}
}

@article{Patil:2026potts,
  title={Learning Potts Models and $ Z\_3 $ Toric Codes: Higher and Ordinary Nishimori Criticality},
  author={Patil, Rushikesh A and P{\"u}tz, Malte and Mukherjee, Rohit and Zhu, Guo-Yi and Trebst, Simon and Ludwig, Andreas WW},
  journal={arXiv preprint arXiv:2608.20268},
  year={2026}
}

@article{Wiese:2026bayesian,
  title={Bayesian phase transition for the critical Ising model: Enlarged replica symmetry in the epsilon expansion and in 2D},
  author={Wiese, Kay Joerg and Das, Alapan and Nahum, Adam},
  journal={arXiv preprint arXiv:2604.23346},
  year={2026}
}

@article{PhysRevB.106.045145,
  title = {Position-dependent excitations and UV/IR mixing in the ${\mathbb{Z}}_{N}$ rank-2 toric code and its low-energy effective field theory},
  author = {Pace, Salvatore D. and Wen, Xiao-Gang},
  journal = {Phys. Rev. B},
  volume = {106},
  issue = {4},
  pages = {045145},
  numpages = {20},
  year = {2022},
  publisher = {American Physical Society},
  doi = {10.1103/PhysRevB.106.045145},
  url = {https://link.aps.org/doi/10.1103/PhysRevB.106.045145}
}

@article{Vijay,
  title = {Fracton topological order, generalized lattice gauge theory, and duality},
  author = {Vijay, Sagar and Haah, Jeongwan and Fu, Liang},
  journal = {Phys. Rev. B},
  volume = {94},
  issue = {23},
  pages = {235157},
  numpages = {9},
  year = {2016},
  month = {Dec},
  publisher = {American Physical Society},
  doi = {10.1103/PhysRevB.94.235157},
  url = {https://link.aps.org/doi/10.1103/PhysRevB.94.235157}
}

@article{Ebisu:2023idd,
    author = "Ebisu, Hiromi and Honda, Masazumi and Nakanishi, Taiichi",
    title = "{Foliated field theories and multipole symmetries}",
    eprint = "2310.06701",
    archivePrefix = "arXiv",
    primaryClass = "cond-mat.str-el",
    reportNumber = "YITP-23-129, RIKEN-iTHEMS-Report-23",
    doi = "10.1103/PhysRevB.109.165112",
    journal = "Phys. Rev. B",
    volume = "109",
    number = "16",
    pages = "165112",
    year = "2024"
}

@article{Pace:2025hpb,
    author = {Pace, Salvatore D. and Aksoy, {\"O}mer M. and Lam, Ho Tat},
    title = "{Spacetime symmetry-enriched SymTFT: from LSM anomalies to modulated symmetries and beyond}",
    eprint = "2507.02036",
    archivePrefix = "arXiv",
    primaryClass = "cond-mat.str-el",
    reportNumber = "MIT-CTP/5884",
    doi = "10.21468/SciPostPhys.20.1.007",
    journal = "SciPost Phys.",
    volume = "20",
    pages = "007",
    year = "2026"
}

@Article{10.21468/SciPostPhys.20.4.117,
	title={{Modulated symmetries from generalized Lieb-Schultz-Mattis anomalies}},
	author={Hiromi Ebisu and Bo Han and Weiguang Cao},
	journal={SciPost Phys.},
	volume={20},
	pages={117},
	year={2026},
	publisher={SciPost},
	doi={10.21468/SciPostPhys.20.4.117},
	url={https://scipost.org/10.21468/SciPostPhys.20.4.117},
}

@misc{Han2026Ordinary, author = {Bo Han and Hiromi Ebisu and Guo-Yi Zhu and Simon Trebst}, title = {Ordinary and Higher Nishimori Structures of the Rank-2 Toric Code}, year = {2026}, note = {to appear} }

@article{Castellani_2005,
   title={Spin-glass theory for pedestrians},
   volume={2005},
   ISSN={1742-5468},
   url={http://doi.org},
   DOI={10.1088/1742-5468/2005/05/p05012},
   number={05},
   journal={Journal of Statistical Mechanics: Theory and Experiment},
   publisher={IOP Publishing},
   author={Castellani, Tommaso and Cavagna, Andrea},
   year={2005},
   month={may},
   pages={P05012--P05012}
}

@article{S_Edwards_1975,
doi = {10.1088/0305-4608/5/5/017},
url = {https://doi.org/10.1088/0305-4608/5/5/017},
year = {1975},
month = {may},
publisher = {},
volume = {5},
number = {5},
pages = {965},
author = {S F Edwards and P W Anderson},
title = {Theory of spin glasses},
journal = {Journal of Physics F: Metal Physics}
}

@article{b3s8-y1wb,
  title = {Measurement-induced phase transitions in quantum inference problems and quantum hidden Markov models},
  author = {P. Kim, Sun Woo and von Keyserlingk, Curt and Lamacraft, Austen},
  journal = {Phys. Rev. Res.},
  volume = {8},
  issue = {2},
  pages = {023155},
  numpages = {42},
  year = {2026},
  month = {May},
  publisher = {American Physical Society},
  doi = {10.1103/b3s8-y1wb},
  url = {https://link.aps.org/doi/10.1103/b3s8-y1wb}
}

@article{PhysRevA.69.032306,
  title = {Anyon computers with smaller groups},
  author = {Mochon, Carlos},
  journal = {Phys. Rev. A},
  volume = {69},
  issue = {3},
  pages = {032306},
  numpages = {26},
  year = {2004},
  month = {Mar},
  publisher = {American Physical Society},
  doi = {10.1103/PhysRevA.69.032306},
  url = {https://link.aps.org/doi/10.1103/PhysRevA.69.032306}
}

@article{Griffiths_1967,
  author = {Griffiths, Robert B.},
  title = {Correlations in Ising Ferromagnets. I},
  journal = {Journal of Mathematical Physics},
  volume = {8},
  number = {3},
  pages = {478--483},
  year = {1967},
  doi = {10.1063/1.1705219}
 }

@article{Kelly_Sherman_1968,
  author = {Kelly, D. G. and Sherman, S.},
  title = {General Griffiths' Inequalities on Correlations in Ising Ferromagnets},
  journal = {Journal of Mathematical Physics},
  volume = {9},
  number = {3},
  pages = {466--484},
  year = {1968},
  month = {March},
  doi = {10.1063/1.1664601}
}

@article{patil2026higher,
  title={Higher Nishimori criticality and exact results at the learning transition of deformed toric codes},
  author={Patil, Rushikesh A and P{\"u}tz, Malte and Trebst, Simon and Zhu, Guo-Yi and Ludwig, Andreas WW},
  journal={arXiv preprint arXiv:2604.06324},
  year={2026}
}
\bibliographystyle{utphys}
\appendix

\section{Griffiths--Kelly--Sherman inequality}\label{gks}

In this appendix, we briefly review the
Griffiths--Kelly--Sherman (GKS) inequalities
\cite{Griffiths_1967,Kelly_Sherman_1968}, which are used in
Sec.~\ref{subsec:exact-correlator-hierarchy} to derive the hierarchy among record averaged correlators.

Consider Ising spins \(\sigma_i=\pm1\) on a finite lattice
\(\Lambda\), with Hamiltonian
\begin{equation}
    H(\bm{\sigma})
    =
    -\sum_{X\subseteq\Lambda}K_X\sigma_X,
    \qquad
    \sigma_X:=\prod_{i\in X}\sigma_i,
    \qquad
    K_X\geq0.
\end{equation}
Here \(X\) denotes an arbitrary subset of lattice sites. The thermal
expectation value of an observable \(F(\bm{\sigma})\) is
\begin{equation}
    \langle F\rangle
    :=
    \frac{1}{Z}
    \sum_{\bm{\sigma}}
    F(\bm{\sigma})
    \exp\left(
        \sum_{X\subseteq\Lambda}K_X\sigma_X
    \right),
\end{equation}
where
\begin{equation}
    Z
    :=
    \sum_{\bm{\sigma}}
    \exp\left(
        \sum_{X\subseteq\Lambda}K_X\sigma_X
    \right).
\end{equation}

The first GKS inequality states that
\begin{equation}
    \langle\sigma_A\rangle\geq0
\end{equation}
for every \(A\subseteq\Lambda\). To see this, expand the Boltzmann
weight as
\begin{align}
    Z\langle\sigma_A\rangle
    &=
    \sum_{\bm{\sigma}}
    \sigma_A
    \prod_{X\subseteq\Lambda}e^{K_X\sigma_X}
   =
    \sum_{\{n_X\geq0\}}
    \left(
        \prod_{X\subseteq\Lambda}
        \frac{K_X^{n_X}}{n_X!}
    \right)
    \sum_{\bm{\sigma}}
    \sigma_A
    \prod_{X\subseteq\Lambda}\sigma_X^{n_X}.
\end{align}
For each choice of \(\{n_X\}\), the spin sum takes the form
\begin{equation}
    \sum_{\bm{\sigma}}\prod_{i\in\Lambda}\sigma_i^{m_i}
    =
    \begin{cases}
        2^{|\Lambda|}, & m_i\ \text{even for every }i,\\
        0,             & \text{otherwise}.
    \end{cases}
\end{equation}
Since every coefficient \(K_X^{n_X}/n_X!\) is nonnegative, every
term in the expansion is nonnegative, proving the first inequality.

The second GKS inequality states that, for arbitrary
\(A,B\subseteq\Lambda\),
\begin{equation}
    \langle\sigma_A\sigma_B\rangle
    -
    \langle\sigma_A\rangle\langle\sigma_B\rangle
    \geq0.
    \label{eq:GKS_second}
\end{equation}
A convenient proof uses two independent copies of the spin
configuration, denoted by \(\bm{\sigma}\) and \(\bm{\tau}\). The
unnormalized connected correlator can be written as
\begin{align}
    &Z^2
    \left(
        \langle\sigma_A\sigma_B\rangle
        -
        \langle\sigma_A\rangle\langle\sigma_B\rangle
    \right)
   =
    \frac{1}{2}
    \sum_{\bm{\sigma},\bm{\tau}}
    (\sigma_A-\tau_A)
    (\sigma_B-\tau_B)
    \exp\left[
        \sum_{X\subseteq\Lambda}
        K_X(\sigma_X+\tau_X)
    \right].
    \label{eq:GKS_doubled}
\end{align}
We introduce variables as
\begin{equation}
    u_i:=\frac{\sigma_i+\tau_i}{2},
    \qquad
    v_i:=\frac{\sigma_i-\tau_i}{2}.
\end{equation}
For any \(C\subseteq\Lambda\), one has
\begin{align}
    \sigma_C+\tau_C
   =
    2
    \sum_{\substack{D\subseteq C\\ |D|\ {\rm even}}}
    u_{C\setminus D}v_D,\quad
     \sigma_C-\tau_C
 =
    2
    \sum_{\substack{D\subseteq C\\ |D|\ {\rm odd}}}
    u_{C\setminus D}v_D,
\end{align}
where
\begin{equation}
    u_D:=\prod_{i\in D}u_i,
    \qquad
    v_D:=\prod_{i\in D}v_i.
\end{equation}
All coefficients in these expansions are nonnegative. Expanding the
exponential in~\eqref{eq:GKS_doubled} in powers of \(K_X\)
therefore expresses its right-hand side as a sum of monomials in
\(u_i\) and \(v_i\) with nonnegative coefficients.

For each site,
\[
    (u_i,v_i)\in
    \{(1,0),(-1,0),(0,1),(0,-1)\}.
\]
The sum of any monomial \(u_i^p v_i^q\) over these four possibilities
is nonnegative: it vanishes if an exponent is odd or if both
\(p,q>0\), and is positive otherwise. Consequently, every term that
survives the sum over \(\bm{\sigma}\) and \(\bm{\tau}\) is
nonnegative, proving~\eqref{eq:GKS_second}.

In the notation of Sec.~\ref{app:generalized-HN}, taking
\(\sigma_A=O_{\bm v}\) and \(\sigma_B=O_{\bm w}\) gives
\begin{equation}
    \langle O_{\bm v}O_{\bm w}\rangle_{\rm clean}
    \geq
    \langle O_{\bm v}\rangle_{\rm clean}
    \langle O_{\bm w}\rangle_{\rm clean}.
\end{equation}
Since
\(
O_{\bm v}O_{\bm w}=O_{\bm v\oplus\bm w},
\)
this is precisely the inequality used to obtain the exact correlator
hierarchy in Theorem~\ref{th}.

\section{Weak measurements of Pauli $X$ in the rank-2 toric code}\label{xpaul}

In this appendix, we consider weak measurements of local Pauli-$X$
operators in the rank-2 toric code, with the wavefunction
deformation also taken in the $X$ basis. In contrast to the
Pauli-$Z$ measurement problem discussed in Sec.~\ref{444}, the
classical Ising variables are now associated with the two
species of $Z$-type stabilizers. We introduce an Ising spin
$s_{x,y}$ on the horizontal link centered at
$(x+\frac12,y)$ and an Ising spin $t_{x,y}$ on the vertical
link centered at $(x,y+\frac12)$:
\begin{equation}
s_{x,y}:=\tau_{(x+\frac12,y)},
\qquad
t_{x,y}:=\tau_{(x,y+\frac12)}.
\label{eq:r2x-spins}
\end{equation}

A local Pauli-$X$ operator gives a classical interaction
consisting of the Ising spins associated with the $Z$-type
stabilizers with which it anticommutes. The three species of
physical qubits therefore give
\begin{align}
b_{1,x,y}(s,t)
&=
t_{x,y-1}t_{x,y},
\label{eq:r2x-b1}
\\
b_{2,x,y}(s,t)
&=
s_{x-1,y}s_{x,y},
\label{eq:r2x-b2}
\\
b_{0,x,y}(s,t)
&=
s_{x,y}s_{x,y+1}t_{x,y}t_{x+1,y}.
\label{eq:r2x-b0}
\end{align}
Here $(x,y)$ in~\eqref{eq:r2x-b0} labels the lower-left
corner of the plaquette hosting the type-0 qubit. Thus,
$X_1$ and~$X_2$ generate two-spin interactions between
neighboring collinear edges, whereas $X_0$ generates a
four-spin interaction surrounding a plaquette.

The associated clean classical model is
\begin{equation}
\begin{split}
H_X(s,t)
=-\sum_{x,y}\big[
K_1t_{x,y-1}t_{x,y}
+K_2s_{x-1,y}s_{x,y}
+K_0s_{x,y}s_{x,y+1}t_{x,y}t_{x+1,y}
\big],
\end{split}
\label{eq:r2x-clean}
\end{equation}
where $K_a\geq0$. The model is invariant under
\begin{equation}
s_{x,y}\longmapsto
\eta_s\eta_D^{\,y}s_{x,y},
\qquad
t_{x,y}\longmapsto
\eta_t\eta_D^{\,x}t_{x,y},
\qquad
\eta_s,\eta_t,\eta_D\in\{\pm1\}.
\label{eq:r2x-symmetry}
\end{equation}
The first two transformations are uniform spin flips, while
the last one is a spatially modulated symmetry. We assume an
infinite lattice or periodic boundary conditions compatible
with~\eqref{eq:r2x-symmetry}.

For Gaussian measurements with strengths
$\Delta_0,\Delta_1,\Delta_2$, the replica theory obtained by averaging over measurement records is
\begin{equation}
-H_R^{(X)}
=
\sum_{a=0,1,2}\sum_q
\left[
K_a\sum_{\alpha=1}^{R}b_{a,q}^{(\alpha)}
+
\frac{\Delta_a}{2}
\sum_{\substack{\alpha,\beta=1\\\alpha\neq\beta}}^{R}
b_{a,q}^{(\alpha)}b_{a,q}^{(\beta)}
\right],
\label{eq:r2x-replica}
\end{equation}
where $q$ runs over the appropriate vertices or plaquettes.
The higher Nishimori manifold is
\begin{equation}
\Delta_a=K_a,
\qquad a=0,1,2.
\label{eq:r2x-HN}
\end{equation}
Since all the interactions in~\eqref{eq:r2x-b1}--\eqref{eq:r2x-b0} are Ising
monomials, the enlarged replica symmetry and the exact
higher Nishimori identities derived in Sec.~\ref{app:generalized-HN} apply without
modification.

There are two spatially modulated loops consisting of Pauli $X$
operators, which we denote by~$W_{D,x}$ and~$W_{D,y}$. Defining \(C\) be the rectangular contour with corners
$(x_L,y_B)$, $(x_R,y_B)$, $(x_R,y_T)$, and
$(x_L,y_T)$ where
\begin{equation*}
      x_L=x_0,\qquad x_R=x_0+2r_x,\qquad
  y_B=y_0,\qquad y_T=y_0+2r_y, 
\end{equation*} 
and all the coordinates being even integers, 
the spatially modulated loops are given by
\begin{eqnarray}
    W_{D,x}=\prod_{i=x_L}^{x_R}\left[X_{2,(i,y_B)}X_{2,(i,y_B+1)}\right]^{i}\times\prod_{i=x_L}^{x_R-1}X_{0,(i+1/2,y_B+1/2)}\times \prod_{j=y_B+1}^{y_T-1}X_{1,(x_R,j)}\notag\\
  \times \prod_{i=x_L}^{x_R}\left[X_{2,(i,y_T-1)}X_{2,(i,y_T)}\right]^{i}\times\prod_{i=x_L}^{x_R-1}X_{0,(i+1/2,y_T-1/2)}\times \prod_{j=y_B+1}^{y_T-1}X_{1,(x_L,j)},\notag\\
       W_{D,y}= \prod_{i=x_L+1}^{x_R-1}X_{2,(i,y_B)}\times \prod_{j=y_B+1}^{y_T-1}\left[X_{1,(x_R-1,j)}X_{1,(x_R,j)}\right]^{j}\times\prod_{j=y_B}^{y_T-1} X_{0,(x_R-1/2,j+1/2)}\notag\\
  \times \prod_{i=x_L+1}^{x_R-1}X_{2,(i,y_T)}\times \prod_{j=y_B+1}^{y_T-1}\left[X_{1,(x_L,j)}X_{1,(x_L+1,j)}\right]^{j}\times\prod_{j=y_B}^{y_T-1} X_{0,(x_L+1/2,j+1/2)}.\label{dxdy}
\end{eqnarray}
Acting translation operators on the modulated loops generates a homogeneous loop. Indeed, 
on the projected sector which does not admit excess electric charges
$G_{\bm r}=+1
~\;\forall\bm r$,
we obtain
\begin{align}
T_xW_{D,x}T_x^{-1}
= W_{D,x}W_H,\quad
T_yW_{D,y}T_y^{-1}
= W_{D,y}W_{H^\prime},
\label{eq:r2x-projected-translations}
\end{align}
where 
\begin{eqnarray}
    W_H=\prod_{i=x_L+1}^{x_R}X_{2,(i,y_B)}X_{2,(i,y_B+1)}\times\prod_{j=y_B+1}^{y_T-1}X_{1,(x_R,j)}X_{1,(x_R+1,j)}\times X_{0,(x_R+1/2,y_B+1/2)}X_{0,(x_R+1/2,y_T-1/2)}\notag\\
    \times\prod_{i=x_L+1}^{x_R}X_{2,(i,y_T-1)}X_{2,(i,y_T)}\times\prod_{j=y_B+1}^{y_T-1}X_{1,(x_L,j)}X_{1,(x_L+1,j)}\times X_{0,(x_L+1/2,y_B+1/2)}X_{0,(x_L+1/2,y_T-1/2)}\notag\\
    W_{H^\prime}=\prod_{i=x_L+1}^{x_R-1}X_{2,(i,y_B)}X_{2,(i,y_B+1)}\times\prod_{j=y_B+1}^{y_T}X_{1,(x_R-1,j)}X_{1,(x_R,j)}\times X_{0,(x_R-1/2,y_B+1/2)}X_{0,(x_R-1/2,y_T+1/2)}\notag\\
    \times\prod_{i=x_L+1}^{x_R-1}X_{2,(i,y_T)}X_{2,(i,y_T+1)}\times\prod_{j=y_B+1}^{y_T}X_{1,(x_L,j)}X_{1,(x_L+1,j)}\times X_{0,(x_L+1/2,y_B+1/2)}X_{0,(x_L+1/2,y_T+1/2)}.
\end{eqnarray}
The loops $W_H$ and $W_{H^\prime}$ are associated with the homogeneous loop. We regard  these two loops as representing the same charge type, corresponding to the fact that there is only a single type of homogeneous magnetic superselection sector described by noncontractible loops (see~\cite{PhysRevB.111.035149}). In contrast to the Pauli-$Z$ sector, where a single dipole loop has two distinct homogeneous descendants, the Pauli-$X$ sector contains two modulated loops whose appropriate lattice translates generate the same homogeneous loop.

Similar to the previous argument (Sec.~\ref{444}), 
the classical images of the open counterparts of the two
modulated loops~\eqref{dxdy} are described by 
\begin{align}
O_{D,x}(\bm r,\bm r')
:=
t_{\bm r}t_{\bm r'},\quad 
O_{D,y}(\bm r,\bm r')
:=
s_{\bm r}s_{\bm r'},
\label{eq:r2x-ODy}
\end{align}
where $\bm r$ and $\bm r'$ are chosen on the same parity
sublattice. Likewise, the classical image of the homogeneous loop reads
\begin{align}
O_H(\bm r,\bm r')
&:=
t_{\bm r}t_{\bm r+\hat{\bm x}}
t_{\bm r'}t_{\bm r'+\hat{\bm x}}=O_{D,x}(\bm r,\bm r')
O_{D,x}(\bm r+\hat{\bm x},
\bm r'+\hat{\bm x})\label{eq:r2x-OHx}
\\
O_{H^{\prime}}(\bm r,\bm r')
&:=
s_{\bm r}s_{\bm r+\hat{\bm y}}
s_{\bm r'}s_{\bm r'+\hat{\bm y}}=O_{D,y}(\bm r,\bm r')
O_{D,y}(\bm r+\hat{\bm y},
\bm r'+\hat{\bm y})
\label{eq:r2x-OHy}
\end{align}
with the relation
\begin{eqnarray}
    O_H(\bm r,\bm r')=b_{0,\bm r}b_{0,\bm r^\prime}O_{H^{\prime}}(\bm r,\bm r').
\end{eqnarray}
To discuss the learnability of the loops, we define the EA-type correlators as
\begin{align}
C_{D,x}(\bm r,\bm r')
:=
\overline{
\left\langle
O_{D,x}(\bm r,\bm r')
\right\rangle_{\bm m}^{\,2}},\quad 
C_{D,y}(\bm r,\bm r'):=
\overline{
\left\langle
O_{D,y}(\bm r,\bm r')
\right\rangle_{\bm m}^{\,2}},\\
C_H(\bm r,\bm r'):=
\overline{
\left\langle
O_H(\bm r,\bm r')
\right\rangle_{\bm m}^{\,2}},\quad 
C_{H^\prime}(\bm r,\bm r')
:=
\overline{
\left\langle
O_{H^\prime}(\bm r,\bm r')
\right\rangle_{\bm m}^{\,2}}.
\label{eq:r2x-correlators}
\end{align}
On the higher Nishimori manifold, these correlators are
equal to the corresponding clean correlation functions. 
Applying the GKS inequality (Appendix.~\ref{gks}) to~\eqref{eq:r2x-OHx} and \eqref{eq:r2x-OHy}, and using
translation invariance, gives
\begin{align}
C_H(\bm r,\bm r')
\geq
\left[C_{D,x}(\bm r,\bm r')\right]^2,
\quad
C_{H^\prime}(\bm r,\bm r')
\geq
\left[C_{D,y}(\bm r,\bm r')\right]^2.
\label{eq:r2x-hierarchy-y}
\end{align}
For $K_0>0$, 
one can further make use of the GKS inequality to find
\begin{eqnarray}
    C_H(\bm r,\bm r')\geq C_{H^\prime}(\bm r,\bm r')\left\langle b_{0,\bm r}b_{0,\bm r^\prime}\right\rangle_{clean}\geq C_{H^\prime}(\bm r,\bm r')\tanh^2 K_0.
\end{eqnarray}
Analogous line of thoughts leads to $ C_{H^\prime}(\bm r,\bm r')\geq C_{H}(\bm r,\bm r')\tanh^2 K_0$. Overall, we have
\begin{eqnarray}
    C_{H}(\bm r,\bm r')\geq \left[C_{D,x}(\bm r,\bm r')\right]^2,\quad C_{H^\prime}(\bm r,\bm r')\geq \tanh^2K_0\left[C_{D,x}(\bm r,\bm r')\right]^2\notag\\
     C_{H^\prime}(\bm r,\bm r')\geq \left[C_{D,y}(\bm r,\bm r')\right]^2,\quad C_{H}(\bm r,\bm r')\geq \tanh^2K_0\left[C_{D,y}(\bm r,\bm r')\right]^2.
\end{eqnarray}
Therefore, long range learnability of either modulated loop
necessarily implies long range learnability of 
homogeneous descendant. The converse does not follow from
these inequalities.
\section{X-cube model}\label{xcube}

In this appendix, we consider the learning problem for the X-cube
stabilizer code. Qubits reside on the links of a cubic
lattice and the Hamiltonian is given by~\cite{Vijay}
\begin{equation}
H_{\rm XC}
=
-\sum_c A_c
-\sum_{v,\mu}B_v^\mu,\label{xc1}
\end{equation}
where
$A_c=\prod_{e\in\partial c}X_e$
is the product over the twelve edges of cube $c$, while
$B_v^\mu$ is the product of four $Z$ operators in the plane
normal to $\mu$ and adjacent to $v$.

\subsection{Weak measurements of Pauli $Z$}
\label{subsec:xcube-height-hierarchy}
We consider local Pauli-$Z$ measurements.
For convenience, we use the dual cubic lattice, in which cube centers
become vertices and edges correspond to plaquettes.
The qubits therefore reside on plaquettes, and the Hamiltonian becomes 
\begin{eqnarray}
    \widehat{H}_{XC}=-\sum_v\widehat{A}_v-\sum_{c,a}\widehat{B}_{c,a}~(a=x,y,z)\label{xca}
\end{eqnarray}
where 
$\widehat{A}_v=\prod_{v\subset p}X_p$
denotes the product over the twelve plaquettes which share the corner $v$.  The other operator is defined as
\begin{eqnarray}
    \widehat{B}_{c,x}:=Z_{c-\hy/2}Z_{c+\hy/2}Z_{c-\hz/2}Z_{c+\hz/2}.
\end{eqnarray}
This operator is the product of $Z$ operators on four plaquettes
on the boundary of the cube.
The operators~$\widehat B_{c,y}$ and $\widehat B_{c,z}$ are defined
by cyclically permuting $x$, $y$, and $z$.

The corresponding classical spin model is then described by
three-dimensional plaquette
Ising model:
\begin{equation}
    P_K(\bm\sigma)=\frac{1}{Z_K}
    \exp\left(\sum_p K_p B_p(\bm\sigma)\right),
    \qquad
    B_p(\bm\sigma)=\prod_{\bm r\in\operatorname{corners}(p)}\sigma_{\bm r},
    \qquad K_p\geq0.
    \label{eq:xcube-clean-plaquette}
\end{equation}
Here \(\bm r=(\hx,\hy,\hz)\) labels a node of a lattice. The distribution is
invariant under independent flips of coordinate planes,
\begin{equation}
    \sigma_{\bm r}\longmapsto
    \eta_x(\hx)\eta_y(\hy)\eta_z(\hz)\sigma_{\bm r},
    \qquad \eta_\mu=\pm1,
    \label{eq:xcube-planar-flips}
\end{equation}
where $\eta_x (x)=-1$ corresponds to a spin flip on a $yz$ plane at fixed $x$. 

For $\hx_0<\hx_1$, $\hy_0<\hy_1$, $\hz_0<\hz_1$, 
the Hamiltonian~\eqref{xca} admits the following loops operator:
\begin{eqnarray*}
    W_z=\prod_{\hz_0\leq \hz\leq \hz_1-1 } \prod_{\hx_0\leq \hx\leq \hx_1 }Z_{(\hx+1/2,\hy_0,\hz+1/2)}Z_{(\hx+1/2,\hy_1,\hz+1/2)}\times \prod_{\hz_0\leq \hz\leq \hz_1-1 } \prod_{\hy_0\leq \hy\leq \hy_1 }Z_{(\hx_0,\hy+1/2,\hz+1/2)}Z_{(\hx_1,\hy+1/2,\hz+1/2)}
\end{eqnarray*}
The classical images of the open counterparts of this loop is given by
\begin{eqnarray}
      O_{z_0,z_1}(\bm u,\bm u')
    =D_{z_0,z_1}(\bm u)D_{z_0,z_1}(\bm u')
   =\sigma_{\bm u,z_0}\sigma_{\bm u,z_1}
      \sigma_{\bm u',z_0}\sigma_{\bm u',z_1}.\label{ew}
\end{eqnarray}
where $\bm u=(\hx,\hy)$ denotes a coordinate of a node on a $xy$ plane.
Physical meaning of this operator is as follows:
Define
\begin{equation}
    D_{z_0,z_1}(\bm u)
    :=\sigma_{\bm u,z_0}\sigma_{\bm u,z_1}.
\end{equation}
Under~\eqref{eq:xcube-planar-flips},
\begin{equation}
    D_{z_0,z_1}(\bm u)\longmapsto
    \eta_z(z_0)\eta_z(z_1)D_{z_0,z_1}(\bm u).
    \label{eq:xcube-dipole-charge}
\end{equation}
Thus this operator is odd under a flip of either endpoint plane,
but invariant under their simultaneous flip and under every plane flip
normal to \(x\) or \(y\). It probes the relative \(\mathbb Z_2\)
symmetry of the two endpoint planes.
 The four spin operator~\eqref{ew} is invariant under all planar subsystem symmetries.
Although \(\langle D_{z_0,z_1}\rangle\) vanishes in a finite volume
symmetry-preserving ensemble, long range order in its two-point function
can diagnose spontaneous breaking of the relative planar symmetry.

We turn to evaluation of the corresponding EA correlator
\begin{equation}
    \mathcal C_{z_0,z_1}(\bm u,\bm u')
    :=\overline{\bigl\langle
        O_{z_0,z_1}(\bm u,\bm u')
    \bigr\rangle_{\bm m}^{\,2}},
    \label{eq:xcube-four-corner-EA}
\end{equation}
where the overline denotes the average over measurement records.
 On the higher Nishimori manifold, \(\Delta_p=K_p\) for every plaquette,
the general identity derived in the main text gives
\begin{equation}
    \mathcal C_{z_0,z_1}(\bm u,\bm u')
    =\bigl\langle O_{z_0,z_1}(\bm u,\bm u')\bigr\rangle_{\rm clean}.
    \label{eq:xcube-four-corner-clean}
\end{equation}
Consequently, its long distance limit diagnoses learnability in the same
relative planar channel as the clean correlation function.

The key observation is that a finite height operator factorizes exactly
into operators supported between intermediate planes. For any sequence
\(z_0<z_1<\cdots<z_M\), we have
\begin{equation}
    D_{z_0,z_M}(\bm u)
    =\prod_{a=0}^{M-1}D_{z_a,z_{a+1}}(\bm u),
    \qquad
    O_{z_0,z_M}(\bm u,\bm u')
    =\prod_{a=0}^{M-1}O_{z_a,z_{a+1}}(\bm u,\bm u').
    \label{eq:xcube-slab-telescoping}
\end{equation}
Every spin on an intermediate plane occurs twice and cancels because
\(\sigma_{\bm r}^2=1\). Equivalently, the intermediate planar symmetry
charges cancel, leaving only the charges at the two endpoint planes.

Applying the GKS inequalities to these Ising monomials, and then using~\eqref{eq:xcube-four-corner-clean}, yields
\begin{align}
    \mathcal C_{z_0,z_M}(\bm u,\bm u')
    &=\left\langle
        \prod_{a=0}^{M-1}O_{z_a,z_{a+1}}(\bm u,\bm u')
      \right\rangle_{\rm clean}
   \geq\prod_{a=0}^{M-1}
       \bigl\langle O_{z_a,z_{a+1}}(\bm u,\bm u')\bigr\rangle_{\rm clean}
   =\prod_{a=0}^{M-1}\mathcal C_{z_a,z_{a+1}}(\bm u,\bm u').
    \label{eq:xcube-partition-hierarchy}
\end{align}
This is the exact hierarchy associated with subdivision in the vertical
direction~\cite{Griffiths_1967,Kelly_Sherman_1968}. It requires neither
statistical independence of the slabs nor a continuous phase transition.
In particular, taking every pair of intermediate planes to be adjacent gives
\begin{equation}
    \boxed{
    \mathcal C_{z,z+h}(\bm u,\bm u')
    \geq\prod_{a=0}^{h-1}
         \mathcal C_{z+a,z+a+1}(\bm u,\bm u')
    }.
    \label{eq:xcube-adjacent-slab-hierarchy}
\end{equation}
The right-hand side is a product of separately record averaged second
moments, rather than a single record average of their product.

For a translation-invariant system and ensemble, let
\(\bm R=\bm u'-\bm u\) and
\(\mathcal C_h(\bm R):=\mathcal C_{z,z+h}(\bm u,\bm u')\).
The relation~\eqref{eq:xcube-adjacent-slab-hierarchy} becomes
\begin{equation}
    \boxed{\mathcal C_h(\bm R)\geq
    [\mathcal C_1(\bm R)]^h}.
    \label{eq:xcube-height-hierarchy}
\end{equation}
Assuming the long distance limits exist, define
\begin{equation}
    q_h:=\lim_{|\bm R|\to\infty}\mathcal C_h(\bm R)
\end{equation}
for fixed $h$. 
It follows that
\begin{equation}
    q_h\geq q_1^h.
\end{equation}
Hence, if the observable involving two adjacent planes remains
learnable at long distances, the same is true for planes
separated by any fixed finite height. Nor does the bound guarantee a nonzero
limit when the height also tends to infinity.
\subsection{Weak measurements of Pauli $X$}
\label{app:xcube-X-hierarchy}

We next consider weak measurements of single-edge $X$ operators, with the
wavefunction deformation also taken in the $X$ basis in the model~\eqref{xc1}. The corresponding
classical model differs from the plaquette Ising model discussed above.
Since at each vertex, $ B_v^x B_v^y B_v^z=1$,
we use $B_v^x$ and $B_v^y$ as local generators and introduce
Ising variables $t_v$ and $s_v$, respectively. Global relations among these
generators give a configuration-independent redundancy in this
parametrization. An $X$ operator on an $x$-directed edge anticommutes with~$B^y$ at its two endpoints, while an $X$ operator on a $y$-directed edge
anticommutes with~$B^x$. On a $z$-directed edge, it anticommutes with both
generators at each endpoint. The associated classical monomials are thus
\begin{equation}
 \mathcal B_{v,x}=s_vs_{v+\hhx},\qquad
 \mathcal B_{v,y}=t_vt_{v+\hhy},\qquad
 \mathcal B_{v,z}=s_vt_vs_{v+\hat z}t_{v+\hat z}.
 \label{eq:xcube-X-monomials}
\end{equation}
Writing $u_v=s_vt_v$, the clean distribution takes the anisotropic
Ashkin--Teller form~\cite{Lyons:2024mapping}
\begin{equation}
 P_K(s,t)=\frac{1}{Z_K}\exp\!\left[
 \sum_v\left(K_xs_vs_{v+\hhx}
 +K_yt_vt_{v+\hhy}+K_zu_vu_{v+\hat z}\right)\right].
 \label{eq:xcube-AT-prior}
\end{equation}
Here $K_\mu\geq0$ are dimensionless deformation parameters. Gaussian
measurements replace the coupling on each edge by
$K_\mu+\Delta_\mu m_{v,\mu}$ in the posterior distribution. Below, we work on the higher Nishimori manifold $\Delta_\mu=K_\mu$.

The classical model has planar subsystem symmetries
\begin{equation}
 \begin{aligned}
 s_{\bm r}\longmapsto\eta_y(\hy)\eta_z(\hz)s_{\bm r},\quad t_{\bm r}\longmapsto\eta_x(\hx)\eta_z(\hz)t_{\bm r},\quad
u_{\bm r}&\longmapsto\eta_x(\hx)\eta_y(\hy)u_{\bm r},
 \end{aligned}
 \label{eq:xcube-AT-symmetry}
\end{equation}
where each $\eta_\mu$ is an arbitrary $\mathbb Z_2$-valued function.
Defining $S^x=s$, $S^y=t$, and $S^z=u$, the classical image of a straight
open $X$ string in direction $\mu$ gives
\begin{equation}
 \prod_{k=0}^{\ell-1}\mathcal B_{v+k\hat\mu,\mu}
 =S_v^\mu S_{v+\ell\hat\mu}^\mu.
 \label{eq:xcube-straight-X-string}
\end{equation}
Its endpoints carry the corresponding lineon excitations.

Consider a rectangular box with positive integer side lengths $a,b,c$
and vertex set
\begin{equation}
 V(a,b,c)=\{0,a\}\times\{0,b\}\times\{0,c\}.
\end{equation}
We define three eight-spin observables on the same vertices as 
\begin{equation}
 Q_x=\prod_{v\in V}s_v,\qquad
 Q_y=\prod_{v\in V}t_v,\qquad
 Q_z=\prod_{v\in V}u_v.
 \label{eq:xcube-box-observables}
\end{equation}
Each is invariant under all the planar subsystem symmetries.
Moreover, $Q_\mu$ is the classical image of the product of the four
open $X$ strings along the $\mu$-directed edges of the box. For example,
\begin{equation}
 Q_x=\prod_{y\in\{0,b\}}\prod_{z\in\{0,c\}}
 \left(s_{0,y,z}s_{a,y,z}\right).
 \label{eq:xcube-four-strings}
\end{equation}
The local identity $u_v=s_vt_v$ implies
\begin{equation}
 Q_xQ_y=Q_z,\qquad Q_yQ_z=Q_x,\qquad Q_zQ_x=Q_y.
 \label{eq:xcube-box-fusion}
\end{equation}
To proceed, we define the EA type correlators as
\begin{equation}
 C_\mu^{(2)}(a,b,c):=
 \overline{\langle Q_\mu\rangle_m^2},\qquad \mu=x,y,z.
 \label{eq:xcube-box-EA}
\end{equation}
On the higher Nishimori manifold, the identity established in Sec.~\ref{app:generalized-HN}
gives $C_\mu^{(2)}=\langle Q_\mu\rangle_{\rm clean}$.
Applying the exact correlator hierarchy to~\eqref{eq:xcube-box-fusion}, we obtain
\begin{equation}
 \boxed{\begin{aligned}
 C_x^{(2)}&\geq C_y^{(2)}C_z^{(2)},\\
 C_y^{(2)}&\geq C_z^{(2)}C_x^{(2)},\\
 C_z^{(2)}&\geq C_x^{(2)}C_y^{(2)}.
 \end{aligned}}
 \label{eq:xcube-box-hierarchy}
\end{equation}
All correlators in each inequality are evaluated on the same box.
These relations hold at finite separation and do not require rotational
symmetry or equality of the couplings $K_x,K_y,K_z$.

To state their long distance consequence, take a sequence of boxes with
$a/\ell\to\alpha_x$, $b/\ell\to\alpha_y$, and
$c/\ell\to\alpha_z$ as $\ell\to\infty$, where all
$\alpha_\mu>0$. If the limits
\begin{equation}
 q_\mu=\lim_{\ell\to\infty}C_\mu^{(2)}(a,b,c)
 \label{eq:xcube-box-order}
\end{equation}
exist, they satisfy $q_x\geq q_yq_z$ and its cyclic permutations.
Thus, long range learning of any two of these observables
implies long range learning of the third. In particular, exactly two of
the three limits cannot be positive while the remaining one vanishes.
This is a constraint on the information carried by products of four
strings; it does not imply that the individual strings are learnable.
The result therefore provides an application of the general hierarchy
to extended observables in a system with planar subsystem symmetry,
without assuming a particular phase diagram.
\section{Evaluation of the correlator}\label{nb}
In this Appendix, we derive~\eqref{eq:qdC-217}. 
To this end, we first establish the group structure of the constraint set.
For each selected constraint $\alpha=(e,N)$, with $e=(i,j)$,
let $\pi_N:G\to G/N$ be the quotient homomorphism.
The constraint is equivalent to $\pi_N(g_i)=\pi_N(g_j)$.
If $g,h\in H_B$, then
\begin{equation}
 \pi_N(g_i h_i)
 =\pi_N(g_i)\pi_N(h_i)
 =\pi_N(g_j)\pi_N(h_j)
 =\pi_N(g_j h_j).
\end{equation}
The constraints are also preserved under componentwise inversion, and the identity configuration satisfies every constraint. Hence $H_B$ is a subgroup of $G^V$.

We rewrite the correlator~\eqref{CBX} as 
\begin{eqnarray}
    C^B_\lambda(x,y)=\frac{1}{d_\lambda|H_B|}\sum_{n\in N_B}
 \sum_{\substack{\boldsymbol g\in H_B\\ g_xg_y^{-1}=n}}   \chi_\lambda(n)\label{a1}
\end{eqnarray}
with $N_B$ being defined as
\begin{equation*}
      N_B(x,y)
 :=\bigl\{g_xg_y^{-1}:\boldsymbol g\in H_B\bigr\}.
\end{equation*}
We have
\begin{eqnarray}
\sum_{\substack{\boldsymbol g\in H_B\\g_xg_y^{-1}=n}}    \chi_\lambda(n)
=\Lambda_n\chi_\lambda(n)
\end{eqnarray}
where $\Lambda_n$ denotes the number of combinations of $\boldsymbol g\in H_B$, satisfying $g_xg_y^{-1}=n$ for given $n$. 
To evaluate~$\Lambda_n$, we introduce the following projection: 
\begin{equation}
 \begin{gathered}
 p_{xy}:H_B\longrightarrow L_B,\qquad
 p_{xy}(\boldsymbol g)=(g_x,g_y),\\
 L_B:=\{(g_x,g_y):\boldsymbol g\in H_B\}\subseteq G\times G.
 \end{gathered}
 \label{eq:qd-A-endpoint-map}
\end{equation}
This is a surjective homomorphism, with kernel
\begin{equation}
 K_{B;x,y}:=\ker p_{xy}
 =\{\boldsymbol g\in H_B:g_x=1_G,\ g_y=1_G\}.
 \label{eq:qd-A-kernel}
\end{equation}
Every constant configuration $g_i=b$ satisfies all selected constraints,
so $(b,b)\in L_B$ for every $b\in G$.
If $(a,b)\in L_B$, multiplication by $(b^{-1},b^{-1})$ on the right
shows that $(ab^{-1},1_G)\in L_B$.
Conversely, if $(n,1_G)\in L_B$, then $(n,1_G)(b,b)=(nb,b)\in L_B$.
Hence, we have
\begin{equation}
 \begin{gathered}
 N_B=\{n\in G:(n,1_G)\in L_B\},\\
 L_B=\{(nb,b):n\in N_B,\ b\in G\}.
 \end{gathered}
 \label{eq:qd-A-LB-NB}
\end{equation}
The first equality makes $N_B$ a subgroup. Also, it is normal as
\begin{equation}
 (h,h)(n,1_G)(h^{-1},h^{-1})=(hnh^{-1},1_G)\in L_B
 \quad(h\in G,\ n\in N_B).
 \label{eq:qd-A-normality}
\end{equation}
Thus $N_B\trianglelefteq G$. Each endpoint pair in $L_B$ has exactly $|K_{B;x,y}|$ preimages under
$p_{xy}$, because each nonempty preimage is a coset of the kernel.
For a fixed $n\in N_B$,~\eqref{eq:qd-A-LB-NB} lists precisely
$|G|$ endpoint pairs $(nb,b)$ with relative value $n$. This leads to
\begin{equation}
 \Lambda_n=|K_{B;x,y}|\,|G|\quad(n\in N_B),
 \qquad |H_B|=|K_{B;x,y}|\,|G|\,|N_B|.
 \label{eq:qd-A-counts}
\end{equation}
Substitution into~\eqref{a1} gives
\begin{equation}
 \begin{aligned}
 C_\lambda^{(B)}(x,y)
 =\frac{|K_{B;x,y}|\,|G|}
 {d_\lambda|K_{B;x,y}|\,|G|\,|N_B|}
 \sum_{n\in N_B}\chi_\lambda(n)=\frac{1}{d_\lambda|N_B|}\sum_{n\in N_B}\chi_\lambda(n).
 \end{aligned}
 \label{eq:qd-A-uniform-average}
\end{equation}
\par
To proceed, we
define the following matrix
\begin{equation}
 P_B^\lambda:=\frac{1}{|N_B|}\sum_{n\in N_B}D^\lambda(n).
 \label{eq:qd-A-projector-definition}
\end{equation}
Since $\chi_\lambda=\operatorname{Tr}D^\lambda$,~\eqref{eq:qd-A-uniform-average} is rewritten as 
$C_\lambda^{(B)}=d_\lambda^{-1}\operatorname{Tr}P_B^\lambda$.
Using closure of the subgroup under multiplication and inversion, we obtain
\begin{equation}
 \begin{aligned}
 (P_B^\lambda)^2
 &=\frac{1}{|N_B|^2}\sum_{n,m\in N_B}D^\lambda(nm)
 =\frac{1}{|N_B|}\sum_{k\in N_B}D^\lambda(k)=P_B^\lambda,\\
 (P_B^\lambda)^\dagger
 &=\frac{1}{|N_B|}\sum_{n\in N_B}D^\lambda(n^{-1})=P_B^\lambda.
 \end{aligned}
 \label{eq:qd-A-projector}
\end{equation}
For the first line, every $k\in N_B$ has exactly $|N_B|$
factorizations $k=nm$. Also $D^\lambda(n)P_B^\lambda=P_B^\lambda$
for all $n\in N_B$. Thus $P_B^\lambda$ is the orthogonal projector
onto vectors fixed by the action of $N_B$.
Further, since $N_B\trianglelefteq G$, one finds
\begin{equation}
 D^\lambda(h)P_B^\lambda D^\lambda(h)^{-1}
 =\frac{1}{|N_B|}\sum_{n\in N_B}D^\lambda(hnh^{-1})
 =P_B^\lambda \qquad(h\in G).
 \label{eq:qd-A-commutation}
\end{equation}
By irreducibility and Schur's lemma, $P_B^\lambda$ is a scalar multiple
of $I_{d_\lambda}$. Its idempotence restricts that scalar to zero or one.
The identity case occurs exactly when every $n\in N_B$ acts as
$I_{d_\lambda}$, namely when $N_B\subseteq\ker D^\lambda$.
Overall, we arrive at 
\begin{equation}
 C_\lambda^{(B)}(x,y)
 =\frac{\operatorname{Tr}P_B^\lambda}{d_\lambda}
 =\begin{cases}
 1,&N_B\subseteq\ker D^\lambda,\\
 0,&\text{otherwise}.
 \end{cases}
 \label{eq:qd-A-zero-one}
\end{equation}
This completes the derivation of~\eqref{eq:qdC-217}.

\section{Derivation of~\eqref{eq:qdC-221}}\label{FKG}
In this appendix, 
%
we prove the relation~\eqref{eq:qdC-221} using positive correlation between increasing
events; Let $2^{\mathcal J}$ be the collection of all subsets
of the constraint label set $\mathcal J$. These subsets are ordered
by inclusion: $B\subseteq D$ means that every constraint selected
in $B$ is also selected in $D$. An event is increasing if it
continues to hold whenever more constraints are selected.
The FKG inequality~\cite{Fortuin1971Correlation} states that a strictly positive probability distribution
on these subsets satisfying
\begin{eqnarray}
     \mu(B\cup D)\mu(B\cap D)\geq\mu(B)\mu(D)
 \qquad\text{for all }B,D\subseteq\mathcal J
\end{eqnarray}
assigns nonnegative correlation to any two increasing events:
\begin{eqnarray}
     \mu(\mathcal A\cap\mathcal B)
 \geq\mu(\mathcal A)\mu(\mathcal B).
\end{eqnarray}
We first verify the condition on $\mu$, and then apply the theorem
to the events $\mathcal A_\lambda$ and $\mathcal A_\eta$.

We consider 
 $B,D\subseteq\mathcal J$. Satisfying all constraints in $B\cup D$
is equivalent to satisfying both sets of constraints.
Moreover, $H_B$ and $H_D$ are subgroups of $H_{B\cap D}$, so their
product set is contained in $H_{B\cap D}$. It follows  that 
\begin{equation}
 H_{B\cup D}=H_B\cap H_D,
 \qquad H_BH_D\subseteq H_{B\cap D}.
 \label{eq:qd-B-subgroups}
\end{equation}
For finite subgroups $H,K$ of a group,
$|HK|=|H||K|/|H\cap K|$.
Indeed, if $hk=h'k'$, then
$t=h^{-1}h'=kk'^{-1}\in H\cap K$; conversely, each such $t$ gives
the same product through $(h',k')=(ht,t^{-1}k)$.
Every element of $HK$ is consequently counted $|H\cap K|$ times.
Applying this identity to~\eqref{eq:qd-B-subgroups} gives
\begin{equation}
 \begin{aligned}
 |H_B|\,|H_D|
 =|H_B\cap H_D|\,|H_BH_D|\leq|H_{B\cup D}|\,|H_{B\cap D}|.
 \end{aligned}
 \label{eq:qd-B-cardinality}
\end{equation}
The other factors in $\mu$ obey the exact identity
\[
 \left(\prod_{\alpha\in B}v_\alpha\right)
 \left(\prod_{\alpha\in D}v_\alpha\right)
 =
 \left(\prod_{\alpha\in B\cup D}v_\alpha\right)
 \left(\prod_{\alpha\in B\cap D}v_\alpha\right).
\]
Every label is counted the same number of times on both sides.
Substituting this identity and~\eqref{eq:qd-B-cardinality} into~\eqref{eq:qdC-215} yields
\begin{equation}
 \mu(B\cup D)\mu(B\cap D)\geq\mu(B)\mu(D).
 \label{eq:qd-B-lattice}
\end{equation}
All labels in $\mathcal J$ have $v_\alpha>0$, and $H_B$ is nonempty
because it contains every constant configuration. Thus $\mu(B)>0$ for every subset $B\subseteq\mathcal J$,
and the required condition on the probability distribution is satisfied.
The events $A_\lambda$ are increasing under the inclusion
of constraint sets. If $B\subseteq D$, then
\begin{equation}
    H_D\subseteq H_B,
    \qquad
    N_D(x,y)\subseteq N_B(x,y).
\end{equation}
Hence $N_B(x,y)\subseteq\ker D^\lambda$ implies
$N_D(x,y)\subseteq\ker D^\lambda$.
Thus $B\in A_\lambda$ implies $D\in A_\lambda$, as required for the application of the FKG inequality.

The FKG inequality~\cite{Fortuin1971Correlation} applied to the
increasing event indicators gives
\begin{equation}
 \mu(\mathcal A_\lambda\cap\mathcal A_\eta)
 \geq\mu(\mathcal A_\lambda)\mu(\mathcal A_\eta).
 \label{eq:qd-B-positive-events}
\end{equation}

Suppose $\nu\prec\lambda\otimes\eta$.
For any $B\in\mathcal A_\lambda\cap\mathcal A_\eta$ and
$n\in N_B$, both $D^\lambda(n)$ and $D^\eta(n)$ are identity matrices.
Their tensor product is therefore the identity on the whole tensor
product space, and in particular on every invariant subspace carrying
$\nu$. Hence $B\in\mathcal A_\nu$, proving
$\mathcal A_\lambda\cap\mathcal A_\eta\subseteq\mathcal A_\nu$.
Therefore, 
\begin{equation}
 \begin{aligned}
 Q_\nu(x,y)
 =\mu(\mathcal A_\nu)\geq\mu(\mathcal A_\lambda\cap\mathcal A_\eta)\geq\mu(\mathcal A_\lambda)\mu(\mathcal A_\eta)=Q_\lambda(x,y)Q_\eta(x,y).
 \end{aligned}
 \label{eq:qd-B-final-chain}
\end{equation}
The first inequality uses the inclusion implied by the fusion rules, and the second uses~\eqref{eq:qd-B-positive-events}, completing the proof of~\eqref{eq:qdC-221}.

\section{ Perturbations around the decoupled higher Nishimori lines}\label{stability}
We examine the stability of the decoupled critical point in the
rank-2 toric code learning problem with Pauli-$Z$ measurements.
When $K_{\square}=\Delta_{\square}=0$, the interactions connect sites
separated by two lattice spacings, and the model decomposes into four
independent parity sublattices, labeled by $\alpha=1,\ldots,4$.
Setting $K_x=K_y=\Delta_x=\Delta_y=K_c$, where $K_c$ is the critical
coupling of the clean square lattice Ising model, places each sector at
its higher Nishimori critical point~\cite{patil2026higher}.

In each sector, the higher Nishimori identity gives
\begin{equation}
 \overline{\langle\sigma_\alpha(\mathbf r)
                   \sigma_\alpha(\mathbf 0)\rangle_m^2}
 =\langle\sigma_\alpha(\mathbf r)
          \sigma_\alpha(\mathbf 0)\rangle_{\mathrm{clean}}
 \sim |\mathbf r|^{-1/4}.
\end{equation}
Thus the replica overlap field
$Q_\alpha^{ab}=\sigma_\alpha^{(a)}\sigma_\alpha^{(b)}$, with $a\ne b$,
has scaling dimension $x_Q=1/8$ in the learning limit $R\to1$.
The single-replica spin field also has $x_\sigma=1/8$, since its
two-point function in this limit is the record averaged first moment,
which equals the clean correlator by the law of total expectation.

Each elementary plaquette contains one site from each parity sublattice.
Its leading continuum operator in replica $a$ is therefore given by
\begin{equation}
 \Phi^{(a)}(\mathbf r)
 =\prod_{\alpha=1}^{4}\sigma_\alpha^{(a)}(\mathbf r),
 \qquad
 X^{ab}(\mathbf r)
 =\Phi^{(a)}(\mathbf r)\Phi^{(b)}(\mathbf r)
 =\prod_{\alpha=1}^{4}Q_\alpha^{ab}(\mathbf r).
\end{equation}
The microscopic displacements within a plaquette contribute derivative
corrections. Independence of the four sectors at the decoupled fixed
point implies
\begin{equation}
 x_\Phi=4x_\sigma=\frac12,
 \qquad x_X=4x_Q=\frac12.
\end{equation}
Here $x_Q$ is the dimension of the overlap within an interacting
learning theory; it is not obtained by adding two independent spin
dimensions. Dimension additivity is used only between the four
decoupled parity sectors.

To leading order, plaquette deformation and Gaussian measurement
produce the perturbation
\begin{equation}
 \delta S=-\int d^2\mathbf r\,
 \left[
  g_K\sum_{a=1}^{R}\Phi^{(a)}(\mathbf r)
  +g_\Delta\sum_{1\le a<b\le R}X^{ab}(\mathbf r)
 \right],
\end{equation}
where $g_K$ and $g_\Delta$ are continuum couplings proportional to
$K_\square$ and $\Delta_\square$, respectively, with nonuniversal
normalization factors absorbed. Independent replica spin flips in each decoupled sector distinguish the components appearing in these sums, so summing over
replica labels does not change their linear scaling dimensions.
Both perturbations consequently have the positive RG eigenvalue
\begin{equation}
 y_\square=2-\frac12=\frac32.
\end{equation}
For a dimensionless bare coupling $g_\square$ along a fixed nonzero
plaquette perturbation direction, the resulting crossover length is given by
\begin{equation}
 \frac{\xi_\times}{a_0}\sim |g_\square|^{-1/y_\square}
 =|g_\square|^{-2/3}.
\end{equation}
Here $a_0$ is a microscopic length. The intra sublattice couplings are kept at $K_x=K_y=\Delta_x=\Delta_y=K_c$, while only $K_\square$ and $\Delta_\square$ are varied.

The product of four critical learning theories is therefore unstable
to plaquette deformation and measurement. Along
$K_\square=\Delta_\square$, the higher Nishimori identities and the
correlator hierarchy remain exact despite this instability.
For unequal couplings, the same power counting establishes departure
from the decoupled fixed point. We leave the subsequent RG flow and the resulting phase diagram for future investigation~\cite{Han2026Ordinary}.

\end{document}